\documentclass[smus]{snow2e}
\usepackage{epsf}
\begin{document}
%
%
%
%
\def \Et {{\rm E}_{\rm T}}
\def \Pt {{\rm P}_{\rm T}}
\def \enu {\epsilon_{\nu}}
\def \stw {$\sin^{2}\theta_{W}$}
\newcommand{\MET}{\mbox{$\protect \raisebox{.3ex}{$\not$}\et$}}
\def\deg{^\circ}
\def\qbar{{\bar q}}
\def\nubar{{\bar \nu}}
\def\W{{\em W\/ }}
\def\Z0{${\em Z^0\/}$}
\def\lum{{\cal L}}
\def\epem{{\rm e^{+}e^{-}}}
\def\tptm{{\tau^{+}\tau^{-}}}
\def\roots{${\sqrt s}\:$}
\def\r#1 {$^{#1}$}
\def\sigW {$\sigma\cdot$B(\W$\rightarrow~$e $\nu$) }
\def\sigZ {$\sigma\cdot$B(\Z0$\rightarrow~\epem$) }
\hyphenation{brem-sstrah-lung proc-ess}
\newcommand{\et}{{\rm E}_{\scriptscriptstyle\rm T}}
\newcommand{\etcone}{{\rm E}_{\scriptscriptstyle\rm T}^{cone}}
\newcommand{\abseta}{\mid \eta^{det} \mid \leq}
\newcommand{\abz}{\mid z \mid \leq}
\newcommand{\fb}{f_{b}}
\newcommand{\ks}{K_{s}^{0}}
\newcommand{\pich}{\pi^{\pm}}
\newcommand{\piz}{ \pi^{0} }
\newcommand{\bigz}{{\cal Z}}
\newcommand{\emf}{f_{em}}
\newcommand{\deltar}{\sqrt{\Delta \eta ^{2}+ \Delta \phi ^{2}}}
\newcommand{\etprime}{{\rm E}_{\scriptscriptstyle\rm T'}}
\newcommand{\ptran}{{\rm P}_{\scriptscriptstyle\rm T}}
\newcommand{\met}{\mbox{$\protect \raisebox{.3ex}{$\not$}\et$}}
\newcommand{\wenu}{W \rightarrow e \nu}
\newcommand{\wmunu}{W \rightarrow \mu \nu}
\newcommand{\wlep}{W \rightarrow \rm{lepton}\, \nu}
\newcommand{\zv}{{\rm z}_{vertex}}
\newcommand{\ppbar}{p\bar{p}}
\newcommand{\qqbar}{q\bar{q}}
\newcommand{\ttbar}{t\bar{t}}
\newcommand{\bbbar}{b\bar{b}}
\newcommand{\ccbar}{c\bar{c}}
\newcommand{\ppbb} { \ppbar \rightarrow  \bbbar }
\newcommand{\zee}{Z \rightarrow e^{+}e^{-} }
\newcommand{\bele}{b \rightarrow c e \nu_{e} }         
\newcommand{\blnu}{b \rightarrow c l \nu_{l} }         
\newcommand{\mtran}{{\rm M}_{\scriptscriptstyle\rm T}}
\newcommand{\acceff}{\rm{A} \times \epsilon}
\newcommand{\geapp}{{\rm\raisebox{-0.7ex}{$\stackrel{\textstyle >}{\sim}$}}}
\newcommand{\leapp}{{\rm\raisebox{-0.7ex}{$\stackrel{\textstyle <}{\sim}$}}}
\def \mc {\multicolumn}
\def \pb    {pb$^{-1} $}
\def \DeltaPhi {$\Delta \phi_{\ell\,\ell \ }$} 
\def \mtop {$M_{top} $}
\def \ztau   {$Z\rightarrow\tau\tau \:$}
\def \DeltaPhil {$\Delta \phi{(\MET,\ell) \ }$} 
\def \DeltaPhij {$\Delta \phi{(\MET,j) \ }$} 
\def \TTbar {$t\overline{t} \; \;$}
\def \dpemu {\Delta \phi_{e\mu}}
\def \Mt {M_{top}}
\def \mtenu  {M_{T}^{e\nu}}
\def \lum {\cal L}
\def \intlum {\int {\cal L} dt}
\def \Zee {Z^{0} \rightarrow e^{+}e^{-}}
\def \Zmumu {Z^{0} \rightarrow \mu^{+}\mu^{-}}
\def \emu {e \mu}  
\def \temux {\ttbar \rightarrow \emu + X}
\def \Ete {E_T^{e}}
\def \Ptmin {P_T^{min}}
\def \Ptmu {P_T^{\mu}}
\def \Etmiss {{\not}{E_T}}
%
\newcommand{\imb}{ \mu {\rm b}^{-1} }
\newcommand{\inb}{ {\rm nb}^{-1} }
\newcommand{\ipb}{ {\rm pb}^{-1} }
\newcommand{\degs}{\mbox{$^{\circ}$}}
\newcommand{\gsim}{\mbox{\small$\stackrel{>}{\sim}$\normalsize}}
\newcommand{\lsim}{\mbox{\small$\stackrel{<}{\sim}$\normalsize}}
%
%
\newcommand{\etal}{{\em et al.}}
\newcommand{\tableskip}{\vskip 5pt plus3pt minus1pt \relax}
\newcommand{\tindent}{\hskip 17pt}
\newcommand{\hfull}{\hspace*{\fill}}
\newcommand{\tline}{\protect\linebreak[4]\hfull}
\newcommand{\linespace}[1]{\protect\renewcommand{\baselinestretch}{#1}
  \footnotesize\normalsize}
%
%
\newcommand{\prl}[1]{Phys. Rev. Lett {\bf #1}}
\newcommand{\prev}[1]{Phys. Rev. {\bf #1}}
\newcommand{\prd}[1]{Phys. Rev. D {\bf #1}}
\newcommand{\zs}[1]{Z. Phys. {\bf #1}}
\newcommand{\ncim}[1]{Nuovo Cim. {\bf #1}}
\newcommand{\plet}[1]{Phys. Lett. {\bf #1}}
\newcommand{\prep}[1]{Phys. Rep. {\bf #1}}
\newcommand{\rmp}[1]{Rev. Mod. Phys. {\bf #1}}
\newcommand{\nphy}[1]{Nucl. Phys. {\bf #1}}
\newcommand{\nim}[1]{Nucl. Instrumen. Meth. {\bf #1}}
\title{Top Quark Physics: Future Measurements}
\author{Raymond~Frey\\ \textit{Department of Physics, University of Oregon,
  Eugene, OR 97403}\vspace*{4pt}\\
David~Gerdes\\ \textit{Department of Physics and Astronomy,
  The Johns Hopkins University, Baltimore, MD 21218}\vspace*{4pt}\\
John~Jaros\\ \textit{Stanford Linear Accelerator Center, Stanford University,
  Stanford, CA 94309}\vspace*{4pt}\\
Steve Vejcik\\ \textit{Department of Physics, University of Michigan, Ann
  Arbor, MI, 48109}\vspace*{4pt}\\ 
Edmond~L.~Berger\\ \textit{Argonne National Laboratory, Argonne, 
  IL 60439}\vspace*{4pt}\\
R.~Sekhar~Chivukula\\ \textit{Department of Physics, Boston University, 
  Boston, MA 02215}\vspace*{4pt}\\
Frank~Cuypers\\ \textit{Paul Scherrer Institute, CH-5232 Villigen PSI,
  Switzerland}\vspace*{4pt}\\
Persis~S.~Drell\\ \textit{Laboratory of Nuclear Studies, Cornell 
  University, Ithaca, NY 14853}\vspace*{4pt}\\
Michael~Fero\\ \textit{Laboratory for Nuclear Science, Massachusetts 
  Institute of Technology, Cambridge, MA 02139}\vspace*{4pt}\\ 
Nicholas~Hadley\\ \textit{Department of Physics, University of Maryland,
  College Park, MD 20742}\vspace*{4pt} \\ 
Tao~Han\\ \textit{Department of Physics, University of California,
  Davis, CA 95616}\vspace*{4pt}\\
Ann~P. Heinson\\ \textit{Department of Physics, University of California,
  Riverside, CA 92521}\vspace*{4pt}\\
Bruce~Knuteson\\ \textit{Department of Physics, Rice University, Houston, 
  TX 77251}\vspace*{4pt}\\
Francisco~Larios\\ \textit{Department of Physics and Astronomy, Michigan State   University, East Lansing, MI 48824}\vspace*{4pt}\\
Hannu~Miettinen\\ \textit{Department of Physics, Rice University, Houston, 
  TX 77251}\vspace*{4pt}\\
Lynne H.~Orr\\ \textit{Department of Physics, University of Rochester,
  Rochester, NY 14627}\vspace*{4pt}\\
Michael E.~Peskin\\ \textit{Stanford Linear Accelerator Center, 
  Stanford University, Stanford, CA 94309}\vspace*{4pt}\\ 
Rajendran~Raja\\ \textit{Fermi National Accelerator Laboratory, Batavia, IL
  60510}\vspace*{4pt}\\
Thomas~Rizzo\\ \textit{Stanford Linear Accelerator Center, 
  Stanford University, Stanford, CA 94309}\vspace*{4pt}\\  
Uri~Sarid\\ \textit{Department of Physics, University of Notre Dame, South
  Bend, IN 46566}\vspace*{4pt}\\  
Carl~Schmidt\\ \textit{Stanford Linear Accelerator Center, 
  Stanford University, Stanford, CA 94309}\vspace*{4pt}\\  
Tim~Stelzer and Zack~Sullivan\\ \textit{Department of Physics, University of
  Illinois, Urbana, IL 61801}\\
       }
\maketitle
\vfill
\vspace*{3in}
\newpage
\mbox{\null}
\newpage
\thispagestyle{plain}\pagestyle{plain}  
\begin{abstract}
We discuss the study of the top quark at future experiments and machines.
Top's large mass makes it a unique probe of physics at the natural electroweak
scale. We emphasize measurements of the top quark's mass, width, and couplings,
as well as searches for rare or nonstandard decays, and discuss the 
complementary roles played by hadron and lepton colliders.
\end{abstract}

\section{Introduction}
\label{sec:intro}
The recent observation of the top quark by the CDF and D0 
collaborations\cite{cdf_obs,d0_obs} has opened up the new field of
top physics. The top quark's measured mass of approximately 
175~GeV\cite{DG_snowmass} is nearly twice the mass of the next most
massive particle, the $Z$ boson. It is also tantalizingly close to 
the natural electroweak scale, set by $v_{Higgs} = 246$~GeV. 
While the Standard Model provides a theoretical context in which the top 
mass can be compared to (and found consistent with) other electroweak data,
it offers no fundamental explanation for the top quark's large mass, which
arises from its large coupling to the symmetry-breaking sector of the theory. 
Precision measurements of the top mass, width, and couplings at future
experiments may therefore lead to a deeper understanding of electroweak 
symmetry-breaking. Such measurements are possible in part because 
the top quark's natural width of 1.4~GeV is much greater
than the hadronization timescale set by $\Lambda_{QCD}$, so that
top is completely described by perturbative QCD. Thus
nature has presented us with the unique opportunity to study the 
weak interactions of a bare quark. It is the conclusion of this
subgroup that precision studies of the top quark should be a high priority
at future machines.

We have concentrated our attention on top physics at the following machines.
The first is the so-called ``TeV-33,'' defined as a luminosity upgrade to 
the Fermilab Tevatron that would result in datasets of 
$\approx$30~fb$^{-1}$ at $\sqrt{s}=2.0$~TeV. For comparison, the goal for 
Tevatron Run II, scheduled to begin in 1999, is 2~fb$^{-1}$ at the 
same energy. We have also
considered the top physics capabilities of the LHC, which will initially 
deliver 10~fb$^{-1}$/year and evolve to 100~fb$^{-1}$/year during 
high-luminosity running. Finally, we have 
considered an $e^+e^-$ linear collider operating
at or above the $\ttbar$ threshold and delivering approximately 
50~fb$^{-1}$/year. We have not explicitly considered a muon collider, 
although its top physics capabilities appear qualitatively similar to those 
of $e^+e^-$ machines provided that detector backgrounds can be controlled. 
We did  not study a ``super $pp$ collider'' in the 60--200~TeV range.
Other recent studies of top physics at the Tevatron can be found
in the TeV2000 report\cite{TeV2000} and references therein, while 
top physics at $e^+e^-$ machines has recently been reviewed by
Murayama and Peskin\cite{Peskin-Murayama} and Frey\cite{Frey96}.

\section{Top Quark Yields} 
At both hadron colliders and lepton colliders, most top quarks are produced
in pairs. 
Each $t$ quark decays immediately to $Wb$, and the observed event topology
depends on the decay mode of the two $W$'s. About 5\% of $\ttbar$ decays are
to the ``dilepton'' final state, which occurs when both $W$'s decay to 
$e\nu$ or $\mu\nu$. The ``lepton+jets'' final state occurs in the 30\%
of $\ttbar$ decays where one $W$ decays into $e\nu$ or $\mu\nu$ and the
other decays into quarks. The remaining 65\% of the decays are to final 
states containing $\tau$ leptons or hadronic jets. In this section we
discuss the yields in these channels at future colliders.

\subsection{Top Yields at Hadron Colliders}

The dominant top quark production mechanism at hadron colliders is 
pair production through $q\bar{q}$ or $gg$ annihilation. The relative
contribution of these two processes at the Tevatron is about 90\%--10\%, while
at the LHC these percentages are reversed. The cross section for top pair
production has been calculated by several authors\cite{xsec_theory}.
For $p\bar{p}$ collisions at the planned Tevatron energy of 
$\sqrt{s}=2.0$~TeV, the 
cross section for $m_t=175$~GeV is calculated to be 7.5~pb, with 
an uncertainty estimated by various groups to be 10-30\%. This is a 40\%
increase over the cross section at 1.8~TeV, and underscores the importance
of even modest upgrades to the Tevatron energy. Thus a 30~fb$^{-1}$ Tevatron
run would result in about 225,000 produced $\ttbar$ pairs.
The LHC ($pp$ collisions
at $\sqrt{s}=14$~TeV) is a veritable top factory, with a calculated $\ttbar$
production cross section of about 760~pb. This would result in about
7.6~million produced $\ttbar$ pairs per experiment in one year of 
low-luminosity LHC running.

In addition, single top 
quarks can be produced through electroweak processes such as $W$-gluon 
fusion or the production of an off-shell $W$ that decays to 
$t\bar{b}$\cite{single}. The single-top production cross section is
about 1/3 the~$\ttbar$ cross section at both the Tevatron and the LHC.
The single-top channels are of particular interest for measurements of 
the top quark width and $V_{tb}$ as described below.

Studies of the top quark at hadron colliders emphasize the dilepton and
lepton+jets decay modes. Because these final states contain isolated 
high-$P_T$ lepton(s)
and missing energy, they are relatively easy to trigger on and reconstruct. 
The dilepton mode has
low backgrounds to begin with, while backgrounds in the lepton+jets
channel can be reduced to an acceptable level by a combination of 
kinematic cuts and $b$-tagging. Recently CDF has demonstrated that
top signals can be identified in the $\tau$ and all-hadronic decay modes
as well, but to establish benchmark yields for future experiments it
is useful to focus on the dilepton and lepton+jets final states. These
yields are obtained from current CDF and D0 acceptances by including
the effects of planned upgrades such as full geometrical coverage for
secondary-vertex $b$-tagging and improved lepton-ID in the region
$1<|\eta|<2.5$\cite{TeV2000}. These acceptances are believed to be 
representative of
any hadron collider detector with charged particle tracking in a 
magnetic field, good lepton identification, and secondary-vertexing
capability.  The assumptions include:
\begin{itemize}
   \item High-$P_T$ charged lepton identifiction with good efficiency
         for $|\eta|<2$
   \item Secondary-vertex $b$-tagging with an efficiency of 50-60\%
         per $b$-jet for  $|\eta|<2$
   \item Ability to tag ``soft leptons'' from $b\rightarrow l\nu X$
         with an efficiency of about 15\% per $b$~jet
   \item Double $b$-tag efficiency of about 40\% per $\ttbar$ event.
         Double-tagged events are a particularly clean sample with low
         combinatoric background and are well-suited for measurement of
         the top mass.
\end{itemize}
Table~\ref{tab:yields_tevatron} shows the expected yields and signal/background
at the Tevatron. The acceptance of the LHC detectors is expected to 
be comparable to that of the Tevatron experiments, so to first order the
yields at the LHC will be greater by a factor equal to the ratio of
the cross sections, approximately 100.
\begin{table}[h]
\begin{center}
\caption{Expected top yields at the Tevatron.}
\label{tab:yields_tevatron}
\begin{tabular}{cccc}
\hline\hline
Mode             & 2~fb$^{-1}$  & 30~fb$^{-1}$ & S/B \\
\hline
Dilepton              &   80 &  1200   & $ 5:1$ \\
$l+\ge3$ jets / 1 $b$ & 1300 & 20,000  & $3:1$ \\
$l+\ge4$ jets / 2 $b$ &  600 &  9000   & $12:1$ \\ 
Single top (all)      &  170 &  2500   & 1:2.2 \\
Single top ($W^*$)    &   20 &   300   & 1:1.3 \\
\hline
\end{tabular}
\end{center}
\end{table}

\subsection{Top Yields at the NLC}
\label{sec:yields-nlc}

The $t\bar{t}$ cross section due to $s$-channel $e^+e^-$
annihilation mediated by $\gamma,Z$ bosons increases abruptly 
at threshold, reaches a maximum roughly 50 GeV
above threshold, then falls roughly as the point cross section
($\sigma_{pt}=87({\rm fb})/s({\rm TeV})$) at higher energy. At $\sqrt{s}=500$
GeV the lowest-order total cross section for unpolarized beams with $m_t=180$ GeV
is $0.54$ pb. The electron beam will be highly polarized ($\sim 90\%$),
and this has a significant effect on $t\bar{t}$ production. The lowest-order
cross section becomes $0.74$ pb ($0.34$ pb) for a fully left-hand 
(right-hand) polarized electron beam.  A design year of
integrated luminosity (50 fb$^{-1}$) at $\sqrt{s}=500$ GeV 
corresponds to roughly $25\times 10^3$  $t\bar{t}$
events. The cross sections for
$t$-channel processes, resulting, for example, in final states such
as $e^+e^-t\bar{t}$ or $\nu\bar{\nu}t\bar{t}$, increase with energy,
but are still relatively small. If it turns out that electroweak
symmetry breaking is strongly coupled, this latter process then turns out
to be of particular interest, as emphasized by 
Barklow\cite{Barklow}.

The $\ttbar$ are produced polarized and, due to initial-state 
bremsstrahlung and gluon radiation, are not always back to back. 
According to expectations, the weak decay $t\rightarrow bW$
proceeds before hadronization can occur. This allows the possibility to perform,
in principle, a complete reconstruction in an environment with little additional
hadronic activity. The rapid top decay also ensures that its spin is transferred
to the $bW$ system, which opens up unique opportunities to probe new physics,
as will be explored in Section \ref{sec:couplings}.
 
The emphasis of most simulations to date has been to 
perform a largely topological event selection, taking advantage
of the multi-jet topology of the roughly $90\%$ of $t\bar{t}$
events with 4 or 6 jets in the final state. Therefore, cuts on
thrust or number of jets drastically reduces the light fermion
pair background. In addition, one can use the multi-jet mass
constraints $M($jet-jet$)\approx M_W$ and
$M($3-jet$)\approx m_t$ for the cases involving
$t\rightarrow bqq^\prime$. Simulation studies\cite{Fujii-1}
have shown that multi-jet resolutions of 5 GeV and
15 GeV for the 2-jet and 3-jet masses, respectively,
are adequate and readily achievable with standard detector resolutions.
A detection efficiency of about 70\% 
with a signal to background ratio of 10 was attained in selecting
6-jet final states just above threshold. These numbers are typical
also for studies which select the 4-jet$+\ell\nu$ decay mode.

Another important technique is that of precision vertex
detection. The present experience
with SLC/SLD can be used as a rather good model of what is possible
at NLC. The small and stable interaction point, 
along with the small beam sizes
and bunch timing, make the NLC ideal for pushing the
techniques of vertex detection. At this meeting, Jackson 
presented\cite{Jackson} simulation results
indicating that $b$-jets can be identified with an efficiency of 60\%
with about 97\% purity.
This has important implications
for top physics. Rather loose $b$-tagging, applied in conjunction
with the standard topological and mass cuts mentioned above,
imply excellent prospects for an efficient and pure top event selection.
Detailed studies employing such a combination of techniques have not
yet been performed, however, and it will be interesting to see what
can be achieved.

The background
due to $W$-pair production is the most difficult to eliminate. 
However, in the limit that the electron beam is fully right-hand
polarized, the $W^+W^-$ cross section is dramatically
reduced.  This allows for experimental control and
measurement of the background. On the other hand, the signal is
also reduced, albeit to a much smaller degree, by running with
right-polarized beam. A possible strategy might be
to run with right-hand polarized beam only long enough to make
a significant check of the component of background due to $W$ pairs.

\section{Mass Measurement at Hadron Colliders}
\label{sec:mass-hadron}
The precision with which the top quark
mass, $m_{t}$, can be measured is an
interesting and important benchmark of
proposed future experiments. Within the Standard Model 
and its extensions $m_{t}$ is a fundamental parameter whose value is
related to the Higgs sector of the
electroweak interaction\cite{peskin}.
 As such, it is desirable to have a measurement with
a precision comparable to that of other electroweak parameters, typically of
the order of $<1\%$.  This would correspond to
an uncertainty of about 2~GeV in $m_{t}$.
Extensions to the Standard Model often predict the value of $m_{t}$, and
a sufficiently precise measurement of $m_{t}$ could also help
distinguish between different models.
For this purpose, it would be of interest to measure the top quark mass
with a precision of about 1~GeV\cite{sarid}.

  The measurements
provided by contemporary experiments at CDF and at D0\cite{tipton,strovink}
have been studied in sufficient detail that the
expected precision at hadron colliders can be 
conservatively extrapolated
with some confidence\cite{TeV2000,heinson,ATLAS}.
Issues relevant to this extrapolation are presented below
as understood from studies of the TeV2000 work but
are believed to be a fair representation of the challenges
for experiments at the LHC as well.
 Other mass-measurement techniques also exist but
have not been explored at the same level of detail.
Control of systematic uncertainties is likely to be the
critical issue in the measurement of $m_{t}$ in any
method.

\subsection{Constrained Fits in Lepton+jets Decays}
\label{sec:lepton-jets}

The most precise direct determination of the
top mass currently comes from reconstructing candidate
top events with a $l\nu{\rm + \ jets}$ topology.  Assuming that
the momenta of all final-state partons except the one neutrino
are measured, that the transverse energy of the system is
conserved, that the $t$ and $\bar{t}$ quarks have a common
mass, and that there are two real $W$ bosons results in an
overconstrained system from which the event kinematics
can be obtained.  The method is of additional interest
because it provides a means of determining other kinematic
features of the $t\bar{t}$ decay such as their transverse
momentum or total invariant mass.

  The accuracy with which
the technique can reconstruct the kinematics is limited by
the ambiguity in making the correspondence between observed
jets and underlying quarks.  Without relying on $b$-tagging,
there are 12 different ways to label the jets as either a
$b$-quark or a light quark from a $W$ and to associate them
with either the $t$ or $\bar{t}$ quark.  If one jet is
$b$-tagged, there are six such combinations and if two jets
are tagged then there are two possibilities.  
Additionally, by requiring the $\nu-{\rm lepton}$
invariant mass to equal $M_W$, the component of the
$\nu$ momentum along the direction of the beam axis can be determined
up to a quadratic ambiguity.  Thus, there are twice as many
kinematically consistent solutions for each event.
By selecting the single solution which best fits the $\ttbar$ hypothesis
according to a $\chi^2$ test,
the reconstruction of the kinematics results
in an estimated top mass for each event.  The measured
top mass is obtained by comparing the event mass distribution
to that predicted by Monte Carlo models for different top
masses using a maximum likelihood method.

Two sources of uncertainty limit the precision with which
this technique can be used to measure $m_{t}$.  The
first is the
statistical uncertainty which arises from the finite detector
resolution and the limited number of events.  Monte Carlo
studies indicate that this source of uncertainty decreases
like $\sigma/\sqrt{N}$.  The intrinsic resolution, $\sigma$
is itself composed of two pieces. The first piece is the 
resolution for those events where the correct assignment is
made between the partons and jets and the second piece is
the resolution for the cases where the incorrect parton-jet
assignment is made.  The relative contribution of each of
these sources varies according to the tagging information
available.  Using no tagging information results in a
resolution dominated by the misassigned component but also
results in the largest number of top events.  Requiring two
tagged jets results in the smallest resolution because of the
much higher fraction of events  with correctly assigned jets but
has a corresponding loss of efficiency.  Table~\ref{tagres} summarizes
the tradeoff in the tagging requirements with the expected 
statistical uncertainty for a luminosity of 2 fb$^{-1}$ at the 
Tevatron or LHC. As shown, the ultimate statistical uncertainty is
a fraction of a GeV for any of the three samples.
\begin{table}[h]
\begin{center}
\caption{Expected statistical precision for measurement of top quark mass 
for differently $b$-tagged subsamples.}
\label{tagres}
\begin{tabular}{lccc}
\hline
\hline
Tags & Number of $t\bar{t}$ Events & Background &
 $\sigma_{m_{t}}$ (GeV)\\ \hline
0 & 20000 & 40000 & 0.3 \\
1 & 12000 & 3000 & 0.3 \\
2 & 4000 & 100 & 0.3 \\
\hline
\hline
\end{tabular}
\end{center}
\end{table}

        The second source of uncertainty in the top mass
measurement is systematic.  The largest
sources of systematic uncertainty arise from differences between the 
observed mass distribution and the prediction from Monte
Carlo and detector simulations.  Such differences 
arise, for instance, in the jet-parton $E_{T}$ scale and in the 
modeling of $t\bar{t}$ production and decay.  Table~\ref{sys} shows
the expected systematic uncertainties for the constrained
fit technique
at future hadron colliders with an integrated luminosity of
${\rm 2 \ fb^{-1}}$.

\begin{table}[h]
\begin{center}
\caption{Expected systematic uncertainties in the measurement
of $m_{t}$ for an integrated luminosity of 10 fb-1 at a hadron
collider.}
\label{sys}
\begin{tabular}{lccc}
\hline
\hline
Systematic&$\sigma_{m_{t}}$ (GeV) \\ \hline
Jet-Parton $E_T$ Scale & 2.0 \\
Event Modeling & 2.0 \\
Background Shape & 0.3 \\
\hline
\hline
\end{tabular}
\end{center}
\end{table}

 Based on present
understanding of $t\bar{t}$ event reconstruction, systematic
uncertainties are expected to limit the ultimate precision
with which the top mass can be measured.  The most important
of these are the precision with which the jet $E_{T}$ scale can
be determined and understanding the multijet environment
of $t\bar{t}$ production.
 
\subsection{Jet $E_{T}$ Scale}
\label{sec:jet-Et}

Jets are 
typically identified using fixed-cone clustering
algorithms.  Monte Carlo models are used to derive
a correspondence between observed jet energies and
the momenta of the underlying partons. An understanding
of the $E_T$ scale therefore involves
both theoretical uncertainties in the model of parton fragmentation to a jet,
and experimental uncertainties in the detector's
measurement of the jet energy.
 
Figure 1 shows a comparison
of energy flow in an annulus about single jets produced
in association with $W$ decays. Comparison of jet anatomies
with this technique between data and Monte Carlo can
be used to quantify theoretical uncertainties in the
jet-parton $E_T$ scale.  Such studies will likely improve
with the size of the control samples and indicate that
theoretical uncertainties in the jet-parton $E_T$ scale can
be managed to better than 1 GeV in future experiments.

\begin{figure}
\begin{center}
\leavevmode
\epsfxsize=3.0in
\epsffile{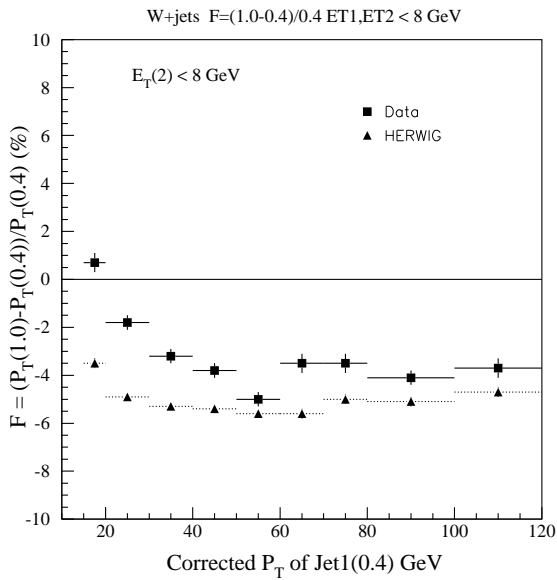}
\caption{Energy flow in annular region around single jets produced in
association with $W$ bosons.}
\end{center}
\label{fig:annulus}
\end{figure}

It is more difficult to reduce the detector effects 
below the present typical value
of 3-4 GeV, and this source of uncertainty could limit the ultimate precision 
of the top mass measurement.  New possibilities for understanding the 
jet-parton $E_T$ scale are offered by control
samples that will be available in future high-statistics data sets.  One 
example is to use
the $W\rightarrow q\bar{q}^{\prime}$ decay in the 
top-quark events themselves to
calibrate the scale.  The dijet mass distribution for 
$W\rightarrow q\bar{q}^{\prime}$
candidates in top events can be compared to a model where
the jet $E_{T}$ scale is varied and used to fit the scale.
A toy Monte Carlo can be used to simulate many such experiments
with the appropriate combinations of signal and background.
Relying on the CDF detector as a model, Fig.~2
shows the distribution of extracted Jet $E_{T}$ scale using the technique
on experiments of varying signal-background composition.
 Each entry in the histogram is the
extracted $E_T$ scale obtained by the method for a single
toy experiment where the true $\et$ scale was perfect.
  The width of the
distribution is the expected precision with which the
Jet $E_T$ scale can be estimated and is seen to be typically
of order 1\%, or a factor of 3-4 better than currently
derived from sample of about $100 {\rm pb^{-1}}$. 
 We therefore conclude that
the jet-parton $E_T$ scale can be controlled to the $\sim$1\%
level, implying a corresponding uncertainty in the
top quark mass of about the same size.

\begin{figure}
\begin{center}
\leavevmode
\epsfxsize=3.0in
\epsffile{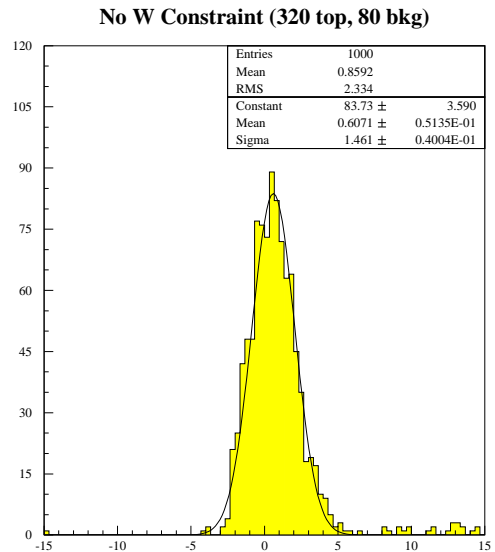}
\caption{Estimated jet-parton $E_T$ scale extracted from toy Monte Carlo
experiments.  Each entry is the measured jet-parton $E_T$
scale obtained from the reconstructed $W\rightarrow q\bar{q}^{\prime}$
decay in top events.  Simulated experiments consisted of samples
of top and background typical for Tevatron Run II data samples.}
\end{center}
\label{fig:wjscale}
\end{figure}
    
\subsection{Uncertainties in Kinematic Modeling}
\label{sec:kine-model}

\begin{figure}
\begin{center}
\leavevmode
\epsfxsize=3.0in
\epsffile{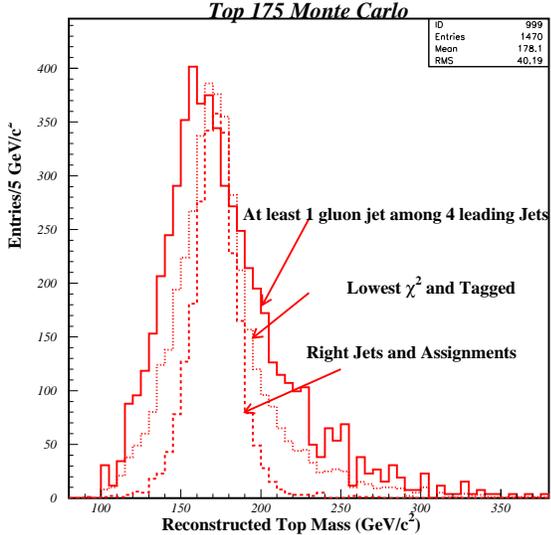}
\caption{Reconstructed mass spectra from $\ttbar$ Monte Carlo 
($m_{t}=175$~GeV) with 
and without the presence of hard gluon radiation.}
\label{fig:hardglue}
\end{center}
\end{figure}

        In addition to the jet-parton $E_T$ scale, uncertainties
 in the top quark mass can arise from the uncertainty in
 modeling the jet environment of top decays.  Constrained fitting techniques
typically associate the leading four jets with the two $b$ jets and two
jets from hadronic $W$ decay; however, initial- or final-state gluon emission
may contaminate the leading four jets with jets that do not arise directly
from $\ttbar$ decay, resulting in a more confused event kinematics.
This effect is modelled by parton shower Monte Carlo programs, such as Herwig.
 Figure~\ref{fig:hardglue}
 shows the invariant mass distribution for top events for those events
 where extra gluon radiation results in a leading jet not
 associated with the partons directly from the $t\bar{t}$ decay.
 Conservatively assuming no information is available on the
 rate of such events implies a corresponding uncertainty on
 the top quark mass of 3~GeV.  This uncertainty is currently
 limited by the lack of a large sample of top quarks with
 which the modeling of jet kinematics can be tested. 
 At the same time, significant theoretical and phenomenological
work has proceeded towards an understanding of gluon radiation
in $t\bar{t}$ events\cite{orr}. In
 datasets with large number of top events, it is evident that
 the understanding of this and other related theoretical issues
 will improve and indeed will be a source of interesting physics
 as well.       

\subsection{Other Mass Measurement Techniques}
\label{sec:other-mass}

        While the constrained fit technique provides the
most precise determination available, other techniques exist, although they 
have not been explored in the same depth.  As described above,
the measurement of the top quark mass can be viewed as simply
comparing a kinematic feature (such as the reconstructed mass)
with that predicted by models for different top masses.  The
same philosophy can be applied, for instance, to underconstrained
topologies such as events where the $t\bar{t}$ decay in the dilepton
mode\cite{raja}.
 This technique is statistically less powerful than the lepton + jets 
method and suffers from similar systematics due to the jet energy scale;
however the
method may complement the more conventional analysis.
Another intriguing possibility is to measure the decay length
of $B$ mesons associated with the $b$-jets in top decays. The
decay length is correlated with the $b$-jet boost and
hence the mass.  It has the additional attractive feature of being a 
mostly tracking-based measurement, and is 
therefore much less dependent on the jet-parton $E_{T}$ scale.
The systematics in this technique, which include uncertainty in
the top quark transverse momentum distribution, need further study.

\subsection{Outlook}
\label{sec:mt-outlook}

It appears that with available technology, the top quark
mass can be measured to a precision of about 1\%, with the caveat that the 
understanding
of theoretical issues dealing with the jet environment in top
decays is thought to be limited primarily by the small number of
events presently available.  It is hoped that systematic
effects from these sources can be brought under control with 
larger samples of data.
  While it is not clear that detector resolution
can be substantially improved, it appears that a program that
relies on control samples in the data can manage the leading
systematic uncertainties to the 1\% level.  The ultimate
resolution as represented by the statistical uncertainty will
be on the order of a few hundred MeV.  The issues of the modeling
of the top kinematics will be crucial but at the same time
will be very interesting tests in and of themselves.
In short, our present understanding of $t\bar{t}$ 
reconstruction at hadron colliders supports the
expectation that the measurement of $m_{t}$
at either the LHC or at an upgraded Tevatron can
be made with the precision thought to be
needed to provide
insights into the Electroweak and Higgs sector of the
Standard Model.

\section{Measurements at the $\ttbar$ Threshold}
\label{sec:threshold}
Production of $\ttbar$ near threshold in $e^+e^-$ (or $\mu^+\mu^-$)
annihilation offers qualitatively unique opportunities for
top physics studies. In addition, in many cases, it promises to allow the
most precise measurements of key parameters. 
The cross section in the $\ttbar$ threshold region depends
sensitively on $m_t$, $\alpha_s$, and $\Gamma_t$, 
and interestingly, also depends
on the top-Higgs Yukawa coupling, $\lambda_t$, and $m_H$.  In this section we
briefly discuss the phenomenology and prospects for these measurements 
near threshold. The $\mu^+\mu^-$ case is not expected to differ significantly
from $e^+e^-$ except for radiative and accelerator effects, and is not 
otherwise specifically discussed.

\subsection{Threshold Shape}
\label{sec:thresh-shape}

\begin{figure}[tb]
\begin{center}
\leavevmode
{\epsfysize=3.0in \epsfbox[50 80 450 450]{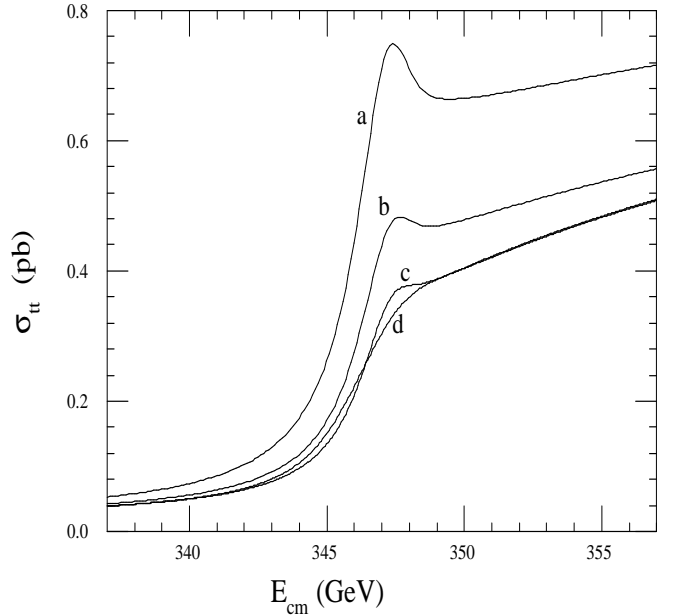}}
\vspace{0.2in}
\caption{ Production cross section for top-quark pairs near 
threshold for $m_t = 175$ GeV.
The theoretical cross section is given by curve (a). The following
energy redistribution effects have been applied to the theory for
the remaining curves: (b) initial-state radiation (ISR); (c): ISR and 
beamstrahlung; (d): ISR, beamstrahlung, and single-beam energy spread.}
\label{fig:thresh}
\end{center}
\end{figure}

In Fig.~\ref{fig:thresh} we show the cross section for $t\bar{t}$ 
production as a function of nominal 
center-of-mass energy $E_{\rm cm}=\sqrt{s}$ for 
$m_t = 175$ GeV. The theoretical cross section, 
indicated as curve (a), is based on the results of 
Strassler and Peskin\cite{Peskin} with $\alpha_s(M_Z^2)=0.120$,
infinite Higgs mass, and nominal Standard Model couplings.  The characterization
of the top threshold is an interesting theoretical issue, and the
theoretical cross section and its associated phenomenology
have been extensively studied\cite{Fadin-1,Peskin,Jezabek-1,
Jezabek-2,Sumino-1,Fujii-1}.
The energy redistribution mechanisms of initial-state radiation, 
beamstrahlung, and single-beam energy spread,  have been 
successively applied to the theoretical curve of Fig.~\ref{fig:thresh}.
Hence, curve (d) includes all effects.  We begin the
discussion of top threshold physics 
with a brief overview of these radiative and accelerator
effects, which are especially important at $\ttbar$ threshold
because of the relatively sharp features in the cross section.

The effects of initial-state radiation (ISR) are appreciable for
high energy electron colliders, where the effective perturbative expansion
parameter for real photon emission, rather than $\alpha/\pi$, is 
$\beta={2\alpha\over\pi}(\ln(s/m_e^2) -1) \approx 1/8$
for $\sqrt{s}=500$ GeV. We use a standard calculation\cite{Fadin-2}
of ISR, which sums the real soft-photon emission to
all orders and calculates the initial state
virtual corrections to second order.
An analytic calculation\cite{Chen} provides
a good approximation for the effects of beamstrahlung at the NLC. 
The figure of merit in the calculation is $\Upsilon = \gamma (B/B_c)$, where 
$\gamma = E_{\rm beam}/m_ec^2$, $B$ is the effective magnetic 
field strength of the beam, and 
$B_c = m_e^2c^3/e\hbar \approx 4\times 10^9$ T. 
When $\Upsilon \ll 1$ the beamstrahlung is in the classical
regime and is readily calculated analytically. 
For example, in the case of the SLAC X-band NLC design, we have
$B\approx 6\times 10^2$ T and
$\Upsilon \approx 0.08$ at $\sqrt{s}=500$ GeV. In this case, there
is an appreciable probability for a beam electron (or positron) to
emit no photons. So the spectrum is well-approximated as
a delta function at $E=E_{\rm cm}$ with a bremsstrahlung-like
tail extending to lower energies. 
The fraction of luminosity within the ``delta
function piece'' of the spectrum resolves the $\ttbar$ threshold structure,
while the remaining luminosity is, for the most part, shifted in energy
well away from threshold. Hence, the primary effect of beamstrahlung is
to reduce the useful luminosity at threshold. The
delta-function fraction of luminosity for
the nominal SLAC X-band NLC design at 500 GeV, for example, is
$43\%$. The energy loss spectrum for initial-state radiation, like
beamstrahlung, has a long tail, and is
also qualitatively similar to beamstrahlung 
in that it is rather likely to have negligibly small energy loss. For example,
$\approx 50\%$ of the total luminosity results in a center
of mass collision energy within $0.1\%$ of the nominal 
$\sqrt{s}\>$\cite{ZDR-IR}.

Hence, to good approximation the combined effect of these
processes is an effective reduction of luminosity at the nominal
$\sqrt{s}$ due to beam particles
which have undergone energy loss $> \Gamma_t$.
We see this in Fig.~\ref{fig:thresh}, although there is clearly
also some smearing out of the threshold shape due to small energy loss.
Of course, there is no control of ISR, except for the choice of beam energy
and accelerated particle---here a muon collider would benefit from the
decreased radiation, where the expansion parameter $\beta$ decreases
from $0.12$ to $0.07$. On the other
hand, the accelerator design will have some
effect on the resulting beamstrahlung spectrum. For example, in changing
$\sqrt{s}$ from 500 GeV to $\ttbar$ threshold, one might choose
to keep the collision point angular divergences constant, 
in which case the spot sizes would increase roughly as $500/350$,
resulting in lower luminosity and decreased beamstrahlung. Alternatively,
scaling the energy at constant beta would result in decreases
only by $\sim\sqrt{500/350}$. So one can expect for the SLAC design
to have the fraction of luminosity unaffected by beamstrahlung (the
delta-function fraction) to be $\sim 50\%$ at $\ttbar$ threshold.

An additional accelerator effect on the threshold shape
results from the energy spread of each
beam in its respective linac. This is the additional effect included in
curve (d) of Fig.~\ref{fig:thresh}, and is characterized by
the FWHM of the energy spread for a single beam, $\Delta E/E$, 
which is a symmetric, non-centrally peaked 
distribution about the nominal beam energy. The calculation of 
Fig.~\ref{fig:thresh} used $\Delta E/E=0.6\%$. This quantity can
be adjusted during operation, typically by $\pm 50\%$,
within some bounds set by the accelerator design.
In Section \ref{sec:scans} we discuss the measurement of the luminosity
spectrum resulting from these effects.

\subsection{Sensitivity to $m_t$ and $\alpha_s$}
\label{sec:mt-alphas.tex}
The threshold enhancement given by the predicted cross section
curve (a) of Fig.~\ref{fig:thresh} reflects the Coulomb-like attraction of the 
produced top pair due to the short-distance QCD potential
\begin{equation}
 V_{\rm QCD}\sim -C_F{\alpha_s(\mu)\over r} \ ,
\label{eq:qcd-thresh}
\end{equation}
where $C_F=4/3$ and 
$\mu$ is evaluated roughly at the scale of the Bohr radius of this 
$t$-$\bar{t}$ bound system:  
$\mu\sim \alpha_s m_t$. This bound state exists,
on average, for approximately one classical revolution before
one of the top quarks undergoes weak decay.  The level spacings of the
QCD potential, approximately given by the Rydberg energy,
$ \sim\alpha_s^2 m_t$, turn out to be comparable to the 
widths of the resonance states, which are 
$\approx 2\Gamma_t$. Therefore the various 
bound states become smeared together, 
where only the bump at the position of the 1S resonance
(at about $347.5$ GeV in Fig.~\ref{fig:thresh}) is
distinguishable. The infrared cutoff imposed by the
large top width also implies\cite{Fadin-1} that the physics is
independent of the long-distance behavior of the QCD potential.
The assumed intermediate-distance potential is also 
found\cite{Fujii-1} to have a negligible impact. Hence, the
threshold physics measurements depend on the short-distance
potential (Eq.~\ref{eq:qcd-thresh}) of perturbative QCD.

An increase of $\alpha_s$ deepens the QCD potential, thereby
increasing the wave function at the origin and producing an
enhanced 1S resonance bump. In addition, the binding energy of
the state varies roughly as the Rydberg energy 
$\sim \alpha_s^2 m_t$. So the larger $\alpha_s$ has the
combined effect of increasing the cross section as well as
shifting the curve to lower energy. The latter effect would also
occur, of course, for a smaller $m_t$. 
Therefore,  measurements of $\alpha_s$ and $m_t$ extracted solely from
a fit to the threshold cross section will be partially correlated, but separable.

In addition to the measurement of the threshold excitation curve,
an interesting and potentially quite useful measurement near threshold
is based upon the observation that the lifetime of the  bound state
is determined by the first top quark to undergo weak decay, 
rather than by annihilation. This implies that the reconstructed
kinetic energy (or momentum) of the top decay products 
reflect the potential energy of the QCD interaction
before decay. Hence, a measurement of the momentum distribution
will be sensitive to $V_{\rm QCD}$ and $\alpha_s$. A larger $\alpha_s$
produces a deeper $V_{\rm QCD}$, hence increasing the kinetic
energy given to the top decay products when the ``spring'' breaks
upon decay of the first of either $t$ or $\bar{t}$.
The theory\cite{Jezabek-1,Sumino-1} and phenomenology\cite{Fujii-1,Hawaii} 
of this physics have been extensively studied. 
The observable used to characterize the distribution is the
peak of the momentum distribution, $p_0$,
which shifts to larger values for larger $\alpha_s$. The best $\sqrt{s}$ 
to run the accelerator for this measurement is about 2 GeV above the 1S peak.
The studies show that $p_0$ is indeed
sensitive to $\alpha_s$. The measurement also has useful sensitivity
to the top width, which arises because a variation in $\Gamma_t$ moves the
average $t$--$\bar{t}$ separation $r_d$ at the time of decay, and hence the
average potential energy $V_{\rm QCD}(r_d)$.

A number of studies have been carried out to simulate measurements
at $t\bar{t}$ threshold. Typically one fixes the width and fits the
threshold shape for the correlated quantities
$m_t$ and $\alpha_s(M_Z^2)$.
For example, a simulation\cite{Fujii-1} 
assuming $m_t=150$ GeV used 1 fb$^{-1}$ for each of 11 scan points. 
If  $m_t$ and $\alpha_s(M_Z^2)$ are left as free parameters, then a simultaneous
2-parameter fit results in errors of $200$ MeV and $0.005$, respectively. 
If one performs a single-parameter fit, holding the other quantity to a fixed
value, the resulting sensitivities approach $100$ MeV and $0.0025$.
An update\cite{Fujii-2} of the 2-parameter fit for $m_t=170$ GeV
gives errors of 350 MeV and $0.007$ for the same 11-point scan.
A similar
simulated scan\cite{Euro-LCWS95} assuming $m_t =180$ GeV and
5 fb$^{-1}$ for each of 10 scan points resulted in single-parameter errors
of $120$ MeV and $0.0025$ for $m_t$ and $\alpha_s$, respectively.
We see that while the error on $m_t$ is remarkably good, the error on
$\alpha_s(M_Z^2)$ is less impressive relative to current measurements.
Of course, it will be very interesting at the outset to compare the threshold
excitation curve with expectations to see, for example, that the $\Delta R$
increase is consistent with the charge and spin of the top quark.
But if the threshold curve can indeed be fit by QCD, then
a reasonable strategy for extracting $m_t$ might be to  
fix $\alpha_s(M_Z^2)$
at the World average value and perform the single-parameter fit of the 
threshold to extract $m_t$. 
The studies cited above have also examined the use of the top momentum
($p_0$) technique. It improves somewhat the precision of the fitted parameters,
typically improving both the $m_t$ and 
$\alpha_s(M_Z^2)$ errors by $\sim 20\%$.
The $p_0$ measurement also has different correlation between mass and
strong coupling than the cross section, hence providing a useful crosscheck.
In fact, Fujii, {\it et al.}~have emphasized that if the scan
energy is referenced to the measured position of the 1S peak, rather
than with respect to $2m_t$ or $\sqrt{s}$,
then the $p_0$ measurement becomes independent of $m_t$. Carried out in
this way, the top momentum measurement would indeed be invaluable
as a crosscheck.
Systematic errors associated with the threshold measurements and 
scan strategies are discussed briefly in Section \ref{sec:scans}.

\subsection{The Top Yukawa Potential}
\label{sec:thresh-yukawa}
\begin{figure}[tbh]
\begin{center}
\leavevmode
{\epsfysize=2.5in \epsfbox[120 280 420 520]{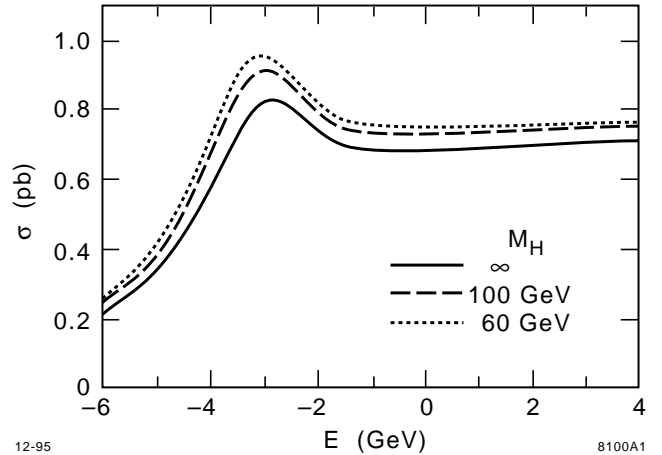}}
\caption{ Cross section near threshold for different Higgs masses
due to the Yukawa potential. $m_t =180$ GeV/c$^2$ was assumed.
The abscissa center-of-mass energy is relative to $2m_t$. }
\label{fig:kuhn-yuk}
\end{center}
\end{figure}

In addition to the QCD potential, the Standard Model predicts
that the $\ttbar$ pair is also
subject to the Yukawa potential associated with Higgs exchange:
\begin{equation}
 V_Y = -{{\lambda_t^2}\over{4\pi}} {{e^{-m_H r}}\over{r}} \ ,
\label{eq:yuk-pot}
\end{equation}
where $m_H$ is the Higgs mass and $\lambda_t$ is the Yukawa
coupling,
\begin{equation}
\lambda_t = \bigl[\sqrt{2}G_F\bigr]^{1/2} \,\beta_H m_t = 
\beta_H m_t /v_{\rm Higgs}
\label{eq:yuk-coupling}
\end{equation}
The dimensionless parameter $\beta_H$ is discussed below.
Because of the extremely short range of the Yukawa potential,
its effect is only on the wave function at the origin,
and hence provides a shift of the cross section across the threshold
region with a slight energy dependence. Fig.~\ref{fig:kuhn-yuk} gives a
calculation\cite{Kuhn-Yukawa} of this effect. It is quite interesting that
because of the large top mass, the Yukawa potential may indeed be observable in
this system. From the various $m_H$ curves given in this calculation,
we clearly see the exponential cutoff of the Yukawa potential
for large $m_H$.

\begin{figure}[tbh]
\begin{center}
\leavevmode
{\epsfysize=2.8in \epsfbox[50 80 450 450]{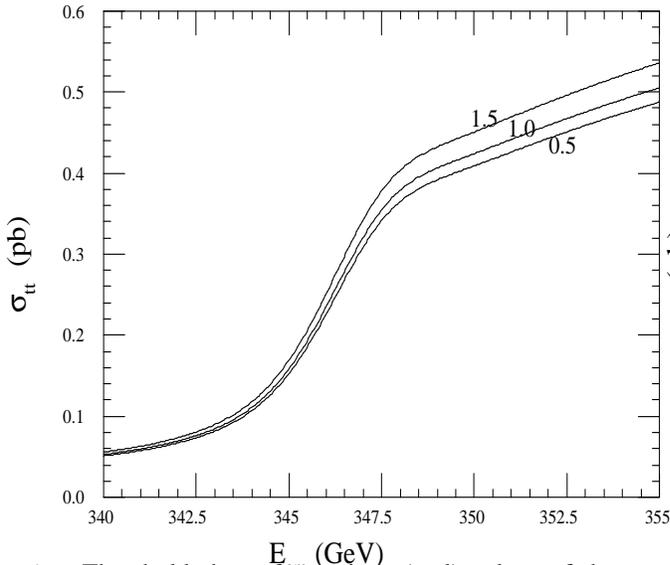}}
\caption{ Threshold shape for various (real) values of the
Yukawa coupling strength $\beta_H$.  All
radiative and beam effects are included, and $m_t=175$ GeV,
$m_H=300$ GeV are used. The different curves corresond to
$\beta_H=1.5$, $1.0$, and $0.5$, as indicated.}
\label{fig:thresh-yuk}
\end{center}
\end{figure}

It is assumed here, of course, that the Higgs bosons(s) will have already been
discovered when such a measurement is undertaken.
However, the Yukawa coupling to fermions is a fundamental
element of electroweak theory, and very likely can only be tested
with top quarks. 
The factor $\beta_H$ in Eq.~\ref{eq:yuk-coupling} is used to
parameterize the strength of the Yukawa coupling and possible 
deviations from the Standard Model, in which $\beta_H=1$.
For example, in two-Higgs-doublet models $\beta_H$ is complex with
real (imaginary) part proportional to $1/\sin\beta$ ($1/\tan\beta$), where
$\tan\beta$ is the usual ratio of Higgs vacuum expectation values.
Hence, these measurements can also be used to help distinguish between
different models of the Higgs sector.
In Section \ref{sec:ttH}~ we review the prospects for the
measurement of $\lambda_t$ in open top production. 
However, the effect of the Higgs field on the $\ttbar$ state at
threshold is unique and it is interesting to see how sensitive
a threshold scan might be. Figure \ref{fig:thresh-yuk}~ shows a calculation
of the cross section across threshold for different values of
$\beta_H$. The values $m_t=175$ GeV and $m_H=300$ GeV were used
and all radiative and accelerator effects are included. (Hence, the $\beta_H=1$
curve corresponds to curve (d) of Fig.~\ref{fig:thresh}.) So one would
have a reasonable sensitivity to this physics with some dedicated running
just above threshold. Fujii\cite{Fujii-2} also applied the previously
mentioned 11 point scan of 1 fb$^{-1}$ per point to the measurement of
$\beta_H$. For larger $m_t$ the accuracy improves, as expected, and
at $m_t=170$ GeV he finds that $\beta_H$ can be measured to $25\%$.

\subsection{The Top Width}
\label{sec:width-nlc}
\begin{figure}[tbh]
\begin{center}
\leavevmode
{\epsfysize=3.0in \epsfbox[70 100 490 480]{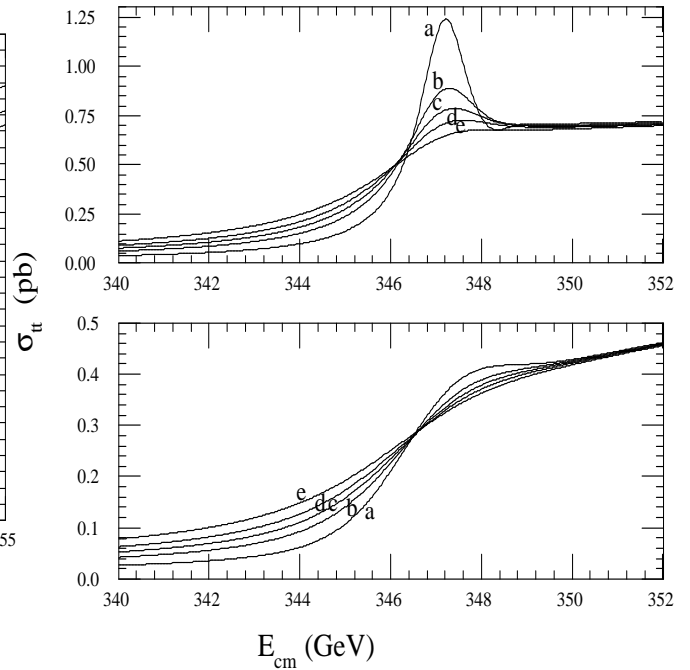}}
\vspace{0.4in}
\caption{ Threshold shape for various values of $\Gamma_t$. The upper
plot is the theoretical prediction, while the lower plot includes all
radiative and beam effects. The different curves corresond to
$\Gamma_t/\Gamma_t^{\rm SM}=$ (a) $0.5$, (b) $0.8$, (c) $1.0$, (d)
$1.2$, and (e) $1.5$. We assumed $m_t=175$ GeV, where
the Standard Model width is $\Gamma_t^{\rm SM}=1.42$ GeV. }
\label{fig:thresh-width}
\end{center}
\end{figure}

Running at $\ttbar$ threshold allows a direct measurement of the
top quark width, $\Gamma_t$, without making any assumptions about
top decay modes. As discussed below in Section \ref{sec:width-had}, this
is especially important for non-standard decays in which top does not
decay to $W$.
On general grounds, we expect the peak cross section of 
a 1S quarkonium bound state to vary with the total width as 
$\Gamma_t^{-1}$, independent of decay modes.  This is shown by the
theoretical curves given in the upper plot of Fig.~\ref{fig:thresh-width}.
After applying ISR and beam effects, the
width is affected as shown in the lower plot of Fig.~\ref{fig:thresh-width}.
In this case, we see that the cross section just below the 1S threshold is
also quite sensitive to the width. The studies cited above indicate sensitivity
to $\Gamma_t$ at the level of 10\% for 50 fb$^{-1}$ of data. However, as
discussed in the next section, any estimate will depend crucially on
the scan strategy employed.

Yet another, quite different observable which is particularly
sensitive to $\Gamma_t$ has been 
studied\cite{Sumino-2,Fujii-1} to help further 
pin down the physics parameters at threshold. The idea is
summarized as follows.
The vector coupling present with $Z$-$t$-$\bar{t}$ and 
$\gamma$-$t$-$\bar{t}$ can proceed to S and D-wave bound
states. On the other hand, the axial-vector coupling present with
$Z$-$t$-$\bar{t}$ gives rise to P-wave states. Hence, 
it is possible to produce interference between S and P-waves
which gives rise to a forward-backward asymmetry 
($A_{FB}$) proportional
to $(v/c)\cos\theta$, where $\theta$ is the usual production polar
angle in the $\ttbar$ rest system. Because of the large width of these
states, due to the large $\Gamma_t$, they do overlap to
a significant extent, and a sizeable $A_{FB}$ develops. The
value of $A_{FB}$ varies from about 5\% to 12\% across the 
threshold, with the minimum value near the 1S resonance.
Since the top width controls the amount of S-P overlap, we expect
the forward-backward asymmetry to be a sensitive method for
measuring $\Gamma_t$. In fact, this has been studied, again by the same
groups as above. Although considerably less sensitive to $\Gamma_t$
than the threshold cross section (about a factor ten in terms of luminosity),
this technique again provides a useful crosscheck of the threshold physics.

\subsection{Systematic Effects and Scan Strategies}
\label{sec:scans}
As indicated in Section \ref{sec:yields-nlc}, an efficient and pure event selection
with good experimental controls appears to be possible. So we expect
the outstanding systematic issue to be the characterization of the redistribution of collision energy  
due to radiation and beam effects, as discussed in
Section \ref{sec:thresh-shape}. This can be quantified by a differential luminosity
spectrum $d{\cal L}/dE$ which describes how the nominal center-of-mass
energy $\sqrt{s}$ is distributed to $e^+e^-$ collision energies $E$. Of course,
this must be determined in order to unfold the physics parameters
from the experiemental scan points. One would hope to measure the luminosity
expended at each scan point to $\sim 1\%$. Fortunately, it is not necessary
to know the radiative and accelerator effects {\it a priori}
at this level of precision. One can, in fact, make an independent measurement
of the luminosity spectrum. As proposed by Frary and Miller\cite{Miller},
the idea is to measure the acollinearity distribution of final state
particles in a $2\rightarrow 2$ process. Bhabha scattering turns out to be
ideal. At intermediate scattering angles 
(about $\theta=20^\circ$ to $\theta=40^\circ$),
Bhabha scattering has a rate $\sim 100$ times that of top production, the
acollinearity can be measured with the requisite accuracy ($<1$ mrad), and it is
theoretically well known at the 1\% level. The acollinearity angle $\theta_A$
for a final-state $e^+e^-$ pair produced at scattering angle $\theta$ is
related to the energy difference $\delta E$ of the initial-state $e^+e^-$ by
$\delta E/E = \theta_A/\sin\theta$, where $E\approx E_{\rm beam}$ 
is their average energy. So starting with the theoretical distribution in
$\theta_A$, one applies contributions due to ISR and beamstrahlung (and 
single-beam energy spread), whose functional forms are known, until
the resulting distribution agrees with the measured one. One then applies
this luminosity spectrum to the top scan data taken over the same
running period.

The other related issue is the determination of the absolute energy scale,
that is, the energy of the beams. This is presently done at the SLC 
using a spectrometer for each spent beam. The
accuracy for $\sqrt{s}$ is 25 MeV ($0.03\%$). Scaling this same error to
top threshold gives 100 MeV accuracy, which is at or below the level of 
error quoted above for a high statistics measurement of $m_t$.
To measure the beam energy, the beams will be briefly taken out of
collision, which eliminates beamstrahlung. (The beamstrahlung-reduced
beam energy measured by the spectrometer is not equivalent to that seen
by collisions since the two sample the beam populations differently.)

\smallskip

Most of the sensitivity to the threshold physics measurements of
$m_t$, $\alpha_s$, $\Gamma_t$, and $\beta_H$ comes from the 
cross section scan across threshold, although as we have seen, the
measurements of top momentum and the forward-backward asymmetry
also provide useful input. These latter two techniques also are more
difficult and demand more study to determine limiting systematics.
Therefore, it is useful to consider how to extract the measurements
solely from the cross section scan. From the discussions above we have
seen that each physics quantity has a different effect on the
threshold shape. So the physics goals will certainly
define the scan strategy.

Expending even a modest fraction of a standard year of luminosity
(50 fb$^{-1}$) at threshold would check the overall physics
of the threshold system and would give an excellent measurement of
the top mass at the level of $\sim 200$ MeV. To concentrate on 
the mass measurement, one would choose to expend luminosity where the
cross section changes most rapidly, at about 346 to 347 GeV for
$m_t=175$ GeV. Assuming a standard model width and fixing
$\alpha_s$ from external measurements, in only $\sim 10$ fb$^{-1}$
one would reach the level of 100 MeV error, 
which is where the systematics
of the absolute energy scale would be expected to become important.

Of course, one would really like to directly measure $\Gamma_t$ given this
opportunity. From Fig.~\ref{fig:thresh-width} we see that measuring
the slope of the threshold rise is required to measure the width. So
one would want to expend luminosity at about 344 and 348 GeV,
as well. Fujii\cite{Fujii-2} finds that 
fixing $\alpha_s$ and performing a 2-parameter
fit to $m_t$ and $\Gamma_t$, the usual $11\times 1$ fb$^{-1}$ scan
gives (statistical) errors of 100 MeV for $m_t$ and
$\delta\Gamma_t /\Gamma_t = 16\%$. If $\Gamma_t$ looked interesting
one could go after the especially sensitive scan energies. Apparently,
the error could be pushed by statistical scaling until the luminosity
systematics become important, at the level of $\sim 1\%$. Hence, a
scan chosen in this way would push the measurement of $\Gamma_t$ to
about 5\% in 50 fb$^{-1}$.

Observing the effect of the Yukawa potential would be unique, and
checking the Yukawa coupling would be a fundamental test. First of
all, one would want to check that the cross section at the 1S
(about $347.5$ GeV for $m_t=175$ GeV) is as expected given the
value of $m_H$ taken from other measurements (see Fig.~\ref{fig:kuhn-yuk}).
This would establish whether the strength of the Yukawa potential
is as expected. Then, from Fig.~\ref{fig:thresh-yuk} we see that one
or two scan points above the 1S would establish the slope and provide
a measurement of $\beta_H$. Again, if the physics demands it, this
measurement could be pushed statistically,
eventually to the level of $\sim 1\%$.

In all cases, a reasonable fraction of the luminosity will have to be
expended just below threshold to measure the background. This fraction
would depend, of course, on the ultimate purity of the event selection,
but 10 to 20\% is a reasonable guess. Since $W^+W^-$ production is 
expected to be the largest background, an important experimental
control is provided by the electron-beam polarization. Flipping between
left and right-handed polarizations would give a huge change in
this background (since the cross section for right-handed production
is tiny) by a predictable amount. So one should expect that the
background fraction can be accurately determined.

In summary, the physics quantities of interest at threshold 
each have different effects on the shape of the threshold curve,
and can be optimally extracted with a cross section scan
employing carefully chosen scan points. In addition, measurements of
the top momentum and forward-backward asymmetry at threshold provide
useful crosschecks of the same quantities.
A modest data set of 10 fb$^{-1}$ would provide a check of the
overall phenomenology and would allow a measurement of $m_t$ with
an error of 100 MeV to 350 MeV, depending upon the scan
and whether $\alpha_s$ is fixed or allowed to be a free parameter.
This luminosity would allow initial measurements of $\alpha_s(M_Z^2)$,
$\Gamma_t$, and the Yukawa coupling $\beta_H$ with errors at
the level of $0.005$, 16\%, and 25\%, respectively. Physics priorities
would push optimization of the scan strategy to concentrate on 
a subset of these quantities, so that with 50 fb$^{-1}$ one could
attain errors of 100 MeV ($m_t$), 0.0025 ($\alpha_s(M_Z^2)$),
5\% ($\Gamma_t$), or 10\% ($\beta_H$). At the current level of
understanding, the measurements become systematics limited near these
errors for $m_t$ and $\alpha_s$, but the width and Yukawa coupling
measurements could be pushed to the level of $\sim 1\%$.

\section{The Top Quark Width and $V_{\lowercase{tb}}$ 
         at Hadron Colliders}
\label{sec:width-had}

In the Standard Model, the top quark decays essentially 100\% of the
time to $Wb$, and the rate for this process leads to a firm prediction
for the top width of $\Gamma_t = 1.4$~GeV (for $m_{t} = 175$~GeV), 
corresponding to a lifetime of $<10^{-25}$s.
A measurement of $\Gamma_t$ is of great interest because $\Gamma_t$ is
affected by any nonstandard decay modes of the top, whether visible or
invisible. Future experiments must therefore address the related questions
``Does top always decay to $Wb$?'' and ``Is $V_{tb}$ equal to 1?''.
That these questions are not equivalent can be seen by considering the
situation with $b$ decays, in which the $b$ quark decays essentially 100\%
of the time to $Wc$ despite the fact that $V_{cb}\approx 0.04$. The 
relatively narrow
width of the $b$ is a consequence of the fact that the quark to which 
it has a large coupling, the top quark, is kinematically inaccessible.
Similarly, a heavy fourth generation quark with a large CKM coupling to top
could allow for a small values of $V_{tb}$ while keeping a large value
of $B(t\rightarrow Wb)$. Thus it is important to measure
$B(t\rightarrow Wb$), $V_{tb}$, and $\Gamma_t$ directly. 

The best measurement of $V_{tb}$ at hadron colliders will come from the
$s$-channel single-top process $q\bar{q}\rightarrow W^*\rightarrow 
t\bar{b}$\cite{stelzer}. These events are detected by requiring a
$W+2$-jet topology where one or both of the jets are $b$-tagged. The
largest background, as in the case of $\ttbar$ events, comes from 
the QCD production of a $W$ in association with one or more $b$-jets.
However, since the single top signal peaks in the 2-jet bin instead of
the 3- and 4-jet bins, this QCD background is considerably higher. 
Nevertheless,
Monte Carlo studies of the signal combined with the observed tagging
rate at CDF in $W$+2-jet events indicate that the signal can be isolated
with a combination of $b$-tagging and kinematic cuts. The expected yield 
for this process is shown in Table~\ref{tab:yields_tevatron}. The advantage
of the $s$-channel single-top process over the higher-rate $t$-channel 
$Wg$ fusion
process is that the cross section can be more reliably calculated
(the uncertainty on the $Q^2$-dependence is only 4\%, as opposed to 
30\% for the $t$-channel
process). The disadvantage of this mode is that has only half the
rate of the $Wg$ single-top process, and therefore requires greater
luminosity. The cross section is proportional to $|V_{tb}|^2$:
\begin{equation}
\label{eqn:sigma_singlet}
     \sigma_{SM} \propto |V_{tb}|^2 B(t\rightarrow Wb).
\end{equation}
Since the branching ratio must be $\le$1, a lower limit on $|V_{tb}|$
is readily obtained from 
\begin{equation}
     |V_{tb}|^2 \ge \sigma_{meas}/\sigma_{SM},
\end{equation}
where $\sigma_{meas}$ is the measured cross section. In 3~fb$^{-1}$ at
the Tevatron, a lower limit of $|V_{tb}|>0.9$ can be obtained, while in
a ``TeV33''-sized sample of 30~fb$^{-1}$ the limit can be extended to 
0.97\cite{stelzer_snowmass}. This measurement will be extremely difficult
at the LHC because the $q\bar{q}$ initial state is swamped by $gg$
contributions. Furthermore the enormous $\ttbar$ cross section at the LHC leads
to significant ``feed-down'' of the $\ttbar$ signal into the 2-jet signal
region.

From Eqn.~\ref{eqn:sigma_singlet}, it is clear that the measurement of the 
single-top production rate via $q\bar{q}\rightarrow W^{*}\rightarrow 
t\bar{b}$ is directly proportional to the partial decay width 
$\Gamma(t\rightarrow Wb)$. In 30~fb$^{-1}$ at the Tevatron, an 8\% measurement
of this partial width should be achievable, where the uncertainty is
likely to be dominated by the 5\% uncertainty on the integrated luminosity. 
To convert this measurement into a measurement
of the total width, it is necessary also to know the branching ratio
$B(t\rightarrow Wb)$. This can be extracted, albeit in a model-dependent
way, from measuring the ratios of branching ratios
$B(t\rightarrow Wb)/B(t\rightarrow Wq)$ and 
$B(t\rightarrow Wq)/B(t\rightarrow ({\rm non-}W+X)$. The first of these
can be  measured
in $\ttbar$ events using the ratio of single to double $b$-tags in the
lepton + jets sample. The requirement of one $b$-tagged jet leaves the 
second $b$-jet unbiased, so that with a known tagging efficiency the 
branching ratio can be measured from the number of additional tags. 
A similar technique can be used in the dilepton sample. Because $b$-tagging
is not required to select high-purity dilepton events, the ratio of
non-tagged to single-tagged events can be used as well. Finally,
one can compare the ratio of double tags in the same jet with two
different tagging techniques (i.e. secondary vertex tags and soft lepton
tags) to double tags in different jets. Small values of 
$B(t\rightarrow Wb)/B(t\rightarrow Wq)$ would result in large values of
this ``same to different jet'' ratio. Measurements of 
$B(t\rightarrow Wb)/B(t\rightarrow Wq)$ using these techniques have already
been performed by CDF\cite{DG_snowmass}, although the current statistical
power is limited. In a 10~fb$^{-1}$ data set, a 1\% measurement of this
ratio appears achievable\cite{TeV2000}. 

This analysis depends on
the model-dependent assumption that the branching ratio of top to
non-$W$ final states is small. For example, if top has a significant
branching ratio to $H^+b$, there will be additional sources of 
$b$-tags from the decays to the charged Higgs, and the above-mentioned analysis
becomes problematic. This is particularly true in the unlucky situation
where $m_{H^+}\approx 80$~GeV, which would give lepton + jets events
kinematically identical to those arising from Standard Model decays of
the $\ttbar$ pair. In the case of a significant branching ratio to $H^+b$,
however, we would expect to under-produce dilepton events, which 
result from two leptonically-decaying $W$'s, relative to lepton + jets
events. This possibility is discussed next.

The ratio $B(t\rightarrow Wq)/B(t\rightarrow {\rm non-}W+X)$ 
can be measured by 
examining the ratio of single-lepton to dilepton events, since number of 
high-$P_T$, isolated charged leptons in the final state counts the 
number of leptonically-decaying $W$'s.
If all $\ttbar$ decays
contain two $W$, the ratio of (produced) single- to dilepton events is 6:1. 
If top can decay to a non-$W$ final state (such as a charged Higgs, or a 
stop quark plus a gaugino) with different branching ratios to leptons, this 
ratio will be modified. Experimentally, top decays to non-$W$ final states
would be indicated by a departure of $\sigma_{DIL}/\sigma_{L+J}$ from
unity, where $\sigma_{DIL}$ and $\sigma_{L+J}$ are the cross sections
measured in the dilepton and lepton+jets modes. Assuming that top always
decays to $Wb$, measurement of this ratio will give a 2\% measurement
of $B(t\rightarrow Wq)/B(t\rightarrow Xb)$ in 30~fb$^{-1}$.
However, if a departure
from the expected value is observed, the interpretation of the results is
model-dependent. For example, the above-mentioned case of
a large branching ration to $H^+b$, with 
$m_{H^+} \approx 80$~GeV, would increase $\sigma_{L+J}$ at the expense 
of $\sigma_{DIL}$.
Of course, such a departure would be evidence for new physics and would
arguably be even more interesting than a measurement of the width.

Combining the measurements of $\Gamma(t\rightarrow Wb)$ from the single
top production cross section, $B(t\rightarrow Wb)/B(t\rightarrow Wq)$ from
the ratios of tags, and $B(t\rightarrow Wq)/B(t\rightarrow Xb)$ from 
the ratio of the dilepton to lepton+jets cross section, a 9\% measurement
of the total width appears achievable with 30~fb$^{-1}$. 

This somewhat
indirect method of obtaining $\Gamma_t$ may be contrasted with the
direct measurement that is possible from a $\ttbar$ threshold scan at the NLC.
Though the two measurements have comparable precision, the approaches
are quite different and illustrate the complementary nature of the two
environments. The $p\bar{p}$ measurement of $\Gamma_t$ relies on collecting 
data from
many different channels (single top, $\ttbar$, with different numbers of
$b$-tags) that span much of the hadron collider top program; it is 
sensitive to a variety of possible sources of new physics. 
But model-dependence may be involved in the interpretation of the result,
especially the measurement of 
$B(t\rightarrow Wq)/B(t\rightarrow {\rm non-}W+X)$. Because the 
model-dependence and sensitivity to new physics are two sides of the
same coin, this may actually be a virtue.
The NLC offers a clean and well-controlled environment where a single
measurement can be performed with high precision and easily interpreted.
Since $\Gamma_t$ will be measured first at hadron colliders,
the $\Gamma_t$ measurement at the NLC will cross-check many aspects 
of the hadron collider program, not just the $\Gamma_t$ measurement
itself.

\section{$V_{\lowercase{tb}}$ at the NLC}
\label{sec:Vtb_nlc}
     The NLC provides a well-understood environment for measuring the
CKM parameter $V_{tb}$.  To date, nearly all our knowledge of this parameter
is inferred from measurements of bottom and strange decays along with the
assumption of the unitarity of the CKM matrix.  Top decays provide the
opportunity to determine $V_{tb}$ directly; with the advent of very large
data sets, they may also allow the measurement of $V_{ts}$.  If the measured
values differ significantly from present expectations, i.e. if $|V_{tb}|\ne 1$
for example, new physics is indicated, perhaps the existence of a new
generation or the violation of weak universality.  These CKM parameters
are also essential for checking the phenomenology of $B$ mixing and
the assumptions underlying CP violation studies in the the $B$ sector.

Just as the $b$ lifetime and the knowledge that $b\rightarrow c$ 
transitions
dominate $b$ decays determine $V_{bc}$, so the top width and the branching
fraction for $t\rightarrow Wb$ fix the partial width 
$\Gamma(t\rightarrow Wb)$, and hence $V_{tb}$.
Explicitly,
\begin{equation}
 \Gamma(t\rightarrow Wb) =  \frac{|V_{tb}|^2 G_F m_t^3\eta_{QCD}}{8\sqrt{2}\pi}
       \left(1-\frac{M_W^2}{m_t^2}\right)^2
       \left(1+\frac{2M_W^2}{m_t^2}\right)
\end{equation}
where $V_{tb}$ scales the universal weak decay rate given by the Fermi coupling
constant, phase space terms, and a QCD correction factor.  To measure the
partial width requires that the total width and the branching fraction to
$Wb$ final states be measured.  The measurement of the total width has
been discussed in Sections~\ref{sec:width-nlc} and~\ref{sec:width-had}.
Studies indicate, for example, that the total width will be measured with an 
error of 5\% given a 50 fb$^{-1}$ scan of the $\ttbar$ threshold.

     What remains is the measurement of the branching fraction, 
$B(t\rightarrow Wb)$.
Although this measurement has not been simulated with full Monte Carlo,
simple arguments can be used to estimate its expected precision.  The
rate of $\ttbar$ production above threshold is well understood theoretically
given the standard model assumptions for top's neutral current couplings.
If one requires six-jet final states, two $b$ jets in the event, dijet
masses consistent with the $W$ mass, and $Wb$ masses consistent with the top
mass, one obtains a clean sample of $\ttbar$ events where both $t$'s have 
decayed
to $Wb$\cite{Fujii-1}. Assuming a net efficiency of 20\% for event 
selection and 
25~fb$^{-1}$ of data 
above $\ttbar$ threshold, there will be about 2000 $\ttbar$ events selected.
The measured cross-section is thus determined to better than 3\% accuracy,
as long as the luminosity is known to the 2\% level or better.  
The branching fraction is then given in terms of the theoretical cross 
section and detection efficiency $\epsilon$ as
\begin{equation}
   B(t\rightarrow Wb)=(\sigma_{meas}/\epsilon\sigma_{SM})^{1/2}.
\end{equation}
The error is most likely dominated by the error in the efficiency.  Assuming
it to be 5\% leads to an uncertainty in the branching fraction of 2.5\%.

     It is likely that a clean and efficient method for tagging a single
top decay in a $\ttbar$ event is possible in the $e^+e^-$ environment.  For
example, one could demand a $b$ jet opposite a hard lepton from a $W$ decay,
and use the measured lepton momentum to test the consistency of the
hypothesis that the $W$ and $b$ jet are back-to-back, as they must be for
top decay near threshold.  Such a single tag lets one measure the branching
fraction directly without assumptions about top quark couplings, simply
by finding the fraction of the remaining top quarks which decay to $Wb$.
Monte Carlo studies are needed to quantify the precision of this
technique.

     The error in the partial width is simply the sum in quadrature of
the errors in the total width and branching fraction, i.e. 5.6\%.  Errors
in the phase space factors and QCD factors are likely small compared to
the error in the partial width, so the error in $V_{tb}$ is about 2.8\%.

     Can one hope to measure $V_{ts}$ or $V_{td}$ at the NLC?  If 
$|V_{ts}|=.04$,
as expected from unitarity constraints on the CKM matrix, 
$B(t\rightarrow Ws)=1.6\times10^{-3}$, leaving a sample of tens of 
events from the a
50 fb$^{-1}$ data set.  Preliminary studies show that requiring a hard 
kaon in the
quark jet, and the absence of secondary $b$ decay vertices, provides strong
rejection against $b$ decay backgrounds.  Even so, substantially more than 
50~fb$^{-1}$ is needed for such a measurement.  The measurement of $V_{td}$ 
is much further out of reach.

\section{Couplings and Form Factors}
\label{sec:couplings}

Due to its rapid weak
decay,  the top spin is transfered directly to the final state
with no hadronization uncertainties, therefore allowing the
helicity dependent information contained in the Lagrangian
to be propagated to the final state. To the extent that
the final state, expected
to be dominated by $bW^+\bar{b}W^-$, can be fully
reconstructed, then a helicity analysis can be performed.
At the NLC or at a muon collider, the top neutral-current couplings are
accessible via the top production vertex. The charged-current couplings
are accessible to both lepton and hadron colliders via top decay.

The study of top couplings, or more generally the interaction form factors,
is broadly speaking an exploration of new physics which is at a very high
energy scale or is otherwise inaccessible directly.  For example, some
models for physics beyond the Standard Model predict new contributions to
dipole moments in top couplings. However, we also know that the Standard
Model itself predicts interesting new behavior for top couplings
and helicity properties. This is due to the very large top mass, making it the
only known fermion with mass near that of $v_{\rm Higgs}=246$ GeV.
The large top Yukawa coupling is an important implication of its unique
connection to electroweak physics. The phenomenology of the top Yukawa
coupling is discussed separately in Sections \ref{sec:ttH}~
and \ref{sec:thresh-yukawa}. Given the important role of longitudinally
polarized $W$ bosons ($W_{\rm long}$) in electroweak symmetry breaking, 
it is interesting that the Standard Model (SM) predicts the fraction of
$W_{\rm long}$
in top decay to be $m_t^2/(m_t^2 + 2M_W^2) = 70\%$, with the remainder
being left-hand polarized. Measuring
this should be a rather straightforward test.

The top neutral-current coupling can be generalized to the following
form for the Z-$t$-$\bar{t}$ or $\gamma$-$t$-$\bar{t}$ vertex factor:
\begin{eqnarray}
{\cal M}^{\mu(\gamma ,Z)} 
=& e\gamma^\mu\left[ Q_V^{\gamma ,Z}F_{1V}^{\gamma ,Z}
+ Q_A^{\gamma ,Z}F_{1A}^{\gamma ,Z}\gamma_5  \right] \nonumber \\
+& {{ie}\over{2m_t}}\sigma^{\mu\nu}k_\nu\left[
Q_V^{\gamma ,Z}F_{2V}^{\gamma ,Z}
+ Q_A^{\gamma ,Z}F_{2A}^{\gamma ,Z}\gamma_5 \right] \ ,
\label{eq:NC}
\end{eqnarray}
which reduces to the familiar SM tree level expression when the form factors are
$ F_{1V}^\gamma = F_{1V}^Z = F_{1A}^Z = 1$, with all others zero. 
The quantities $Q_{A,V}^{\gamma ,Z}$ are the usual SM coupling constants:
$Q_V^{\gamma}=Q_A^{\gamma}={2\over 3}$, 
$Q_V^Z = (1-{8\over 3}\sin^2\theta_W)/(4\sin\theta_W\cos\theta_W)$,
and $Q_A^Z = -1/(4\sin\theta_W\cos\theta_W)$.
The non-standard couplings $F_{2V}^{\gamma ,Z}$ and $F_{2A}^{\gamma ,Z}$
correspond to electroweak magnetic and electric dipole moments, respectively. 
While these couplings are zero at 
tree level in the SM, the analog of
the magnetic dipole coupling is expected to attain a value 
$\sim\alpha_s/\pi$ due to corrections
beyond leading order. On the other hand, the electric dipole term violates
CP and is expected to be zero in the SM through two
loops~\cite{Suzuki}. Such
a non-standard coupling necessarily involves a top spin flip, 
hence is proportional to $m_t$. In fact, many extensions of the Standard 
Model\cite{Bernreuther,Soni-1} involve CP violating phases which give
rise to a top dipole moment of ${\cal O}(10^{-21})$ e-m
at one loop, which is about
ten orders of magnitude greater than the SM expectation, and
may be within the reach of future experiments, as discussed below. 
A study of anomalous chromomagnetic moments was presented\cite{Rizzo}
at this meeting using the gluon energy distribution in $t\bar{t}g$
events, which was also found to be sensitive to the electroweak
neutral-current couplings.

For the top charged-current coupling we can write the 
$W$-$t$-$b$ vertex factor as
\begin{eqnarray}
{\cal M}^{\mu,W} 
=&{g\over\sqrt{2}}\gamma^\mu\left[ P_L F_{1L}^W 
+  P_R F_{1R}^W \right] \nonumber \\
+& {{ig}\over{2\sqrt{2}\,m_t}}\sigma^{\mu\nu}k_\nu\left[
 P_L F_{2L}^W + P_R F_{2R}^W \right] \ ,
\label{eq:CC}
\end{eqnarray}
where the quantities $P_{L,R}$ are the left-right projectors.
In the SM we have $F_{1L}^W = 1$ and all others zero.
The form factor $F_{1R}^W$
represents a right-handed, or $V+A$, charged current component.

\subsection{Helicity Analysis at NLC}
\label{sec:helicity-nlc}

\begin{figure}[tbh]
\begin{center}
\leavevmode
\epsfysize=2.0in \epsfbox{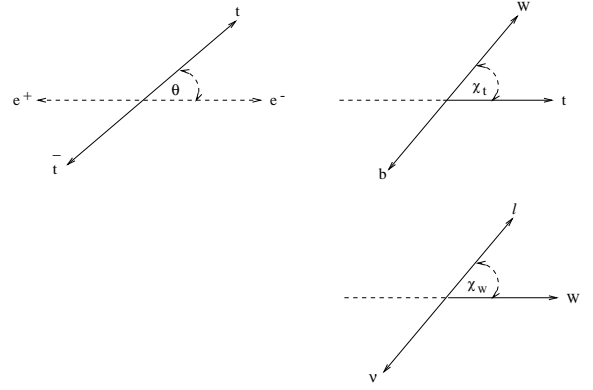}
\end{center}
\caption{Definitions of helicity angles. (a) Production angle $\theta$ in
$t\bar{t}$ proper frame; (b) $\chi_t$ measured in the top proper frame as shown;
and (c) $\chi_W$ in the W proper frame.}
\label{fig:helicity-angles}
\end{figure}

Top pair production above threshold at NLC (or a muon collider)
will provide a unique opportunity to measure
simultaneously all of the top charged and neutral-current couplings. 
In terms of helicity amplitudes, the form factors obey distinct dependences
on the helicity state of $e^-$, $e^+$, $t$, and $\bar{t}$, which can be
accessed experimentally by beam polarization and the measurement of the
decay angles in the final state. These helicity angles can be defined as shown
in Fig.~\ref{fig:helicity-angles}. The angle $\chi_W$ is defined in the 
W proper frame, where the
W direction represents its momentum vector in the limit of zero magnitude.
The analgous statement holds for the definition of $\chi_t$.
As mentioned earlier, the case where the W is longitudinally
polarized is particularly relevant for heavy top, and the 
$\chi_t$ and $\chi_W$ distributions are sensitive to this
behavior. 
Experimentally, all such angles, including the angles corresponding
to $\chi_t$ and $\chi_W$ for the $\bar{t}$ hemisphere, 
are accessible. Given the large number of constraints available
in these events, full event reconstruction is entirely feasible.
To reconstruct $\theta$ one must also take into
account photon and gluon radiation. Photon radiation from the initial state is an important effect, which,
however, represents a purely longitudinal boost which can be 
handled\cite{Yuan} within the 
framework of final-state mass constraints.
Gluon radiation can be more subtle.
Jets remaining after reconstruction of $t$ and $\bar{t}$ can be due to gluon
radiation from $t$ or $b$, and the correct assignment must be decided based on
the kinematic constraints and the expectations of QCD. 

\begin{figure}[tb]
\begin{center}
\leavevmode
\epsfxsize=2.3in \epsfbox[200 320 420 700]{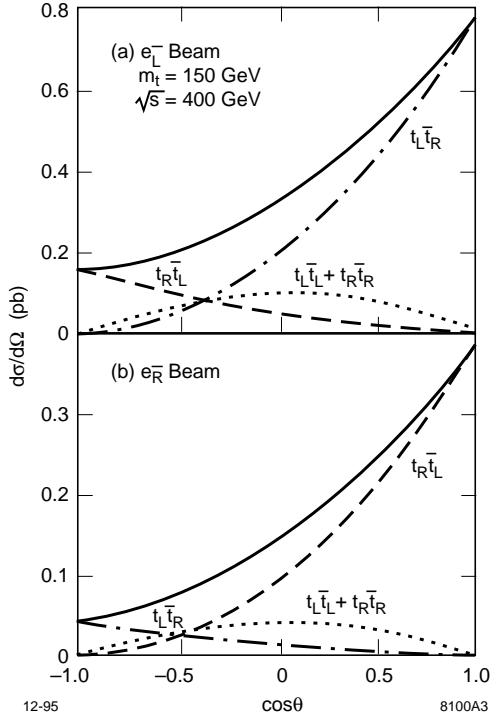}
\end{center}
\caption{
Production angle for $t\bar{t}$ for the possible final-state helicity
combinations, as indicated, for (a) left-polarized electrons, and (b)
right-polarized electrons. The complete cross sections are the solid curves.}
\label{fig:prod-angles}
\end{figure}

The distributions of the production angle $\theta$ 
for the SM in terms of the various helicity states are 
given\cite{Peskin-2} in Fig.~\ref{fig:prod-angles} 
for left and right-hand polarized electron beam.
We see, for example, that for left-hand polarized electron beam, top quarks
produced at forward angles are predominantly left handed, while 
forward-produced top quarks
are predominantly right handed when the electron beam is right-hand polarized.
These helicity amplitudes combine to produce the following general form
for the angular distribution~\cite{Yuan}:
\begin{equation}
{d\sigma\over{d\cos\theta}} = 
{\beta_t\over{32\pi s}}\left[
c_0\sin^2\theta + c_+(1+\cos\theta)^2 + c_-(1-\cos\theta)^2 \right]
\end{equation}
where $c_0$ and $c_\pm$ are functions of the form factors
of Eq.~\ref{eq:NC}, including any non-standard couplings.
The helicity structure of the event is highly constrained by
the measurements of
beam polarization and production angle. 
An alternative analysis framework has been proposed\cite{Parke} involving
a beam-axis system, which might provide higher purity if the final states can 
be only partially reconstructed.

We now outline an analysis\cite{Frey96,Fero}
to measure or set limits on the complete set of form factors defined
in Eqs.~\ref{eq:NC} and \ref{eq:CC}.
We consider a modest integrated luminosity of 10 fb$^{-1}$, 
$m_t = 180$ GeV, and $\sqrt{s} = 500$ GeV.
Electron beam polarization is assumed
to be $\pm 80\%$. 
The decays are assumed to be $t\rightarrow bW$.
In general, one needs to distinguish $t$ from $\bar{t}$. The most
straightforward method for this is to demand that at least one of the W
decays be leptonic, and to use the charge of the lepton as the tag.
(One might imagine using other techniques, for example with topological 
secondary vertex detection one could perhaps distinguish $b$ from $\bar{b}$.)  So we assume the following decay chain: 
\begin{equation}
t\bar{t}\rightarrow b\bar{b}WW \rightarrow 
b\bar{b}q\bar{q}^\prime\ell\nu ,
\end{equation}
where $\ell=e,\mu$. The branching fraction for this decay chain is 
$8/27$. 

Now, since the top production and decay information is correlated, it
is possible to combine all relevant observables to ensure maximum sensitivity
to the couplings. In this study, a likelihood function is used to combine
the observables. We use the Monte Carlo generator
developed by Schmidt\cite{Schmidt}, which includes 
$t\bar{t}(g)$ production to ${\cal O}(\alpha_s)$.  
Most significantly, the Monte Carlo correctly
includes the helicity information at all stages.
The top decay products, including any jets due to 
hard gluon radiation, must be correctly assigned with good probability. 
The correct assignments are rather
easily arbitrated using the $W$ and top mass constraints. 
When the effects of initial-state radiation and 
beamstrahlung are included, it has been shown\cite{Yuan} that
the correct event reconstruction can be performed
with an efficiency of about 70\%. The overall
efficiency of the analysis, including branching fractions,
reconstruction efficiency, and acceptance, is about 18\%.

After simple, phenomenological detection resolution
and acceptance functions are applied, the resulting helicity 
angles (see Fig.~\ref{fig:helicity-angles}) are then used 
to form a likelihood which is the square of the theoretical 
amplitude for these angles 
given an assumed set of 
form factors. 
Table~\ref{table:top-couplings} 
summarizes some of the results of this analysis. 
We see that even
with a modest integrated luminosity of 10 fb$^{-1}$ at $\sqrt{s}=500$ GeV,
the sensitivity to the form factors is quite good, at the level of
5--10\% relative to SM couplings. In terms of real units, 
the 90\% CL limits for $F_{2A}^Z$ of $\pm 0.15$, for example, 
correspond to a $t$-$Z$ electric dipole moment of 
$\sim 8\times 10^{-20}$ e-m. Other studies\cite{Soni-2,Yuan,Cuypers}
have found similar sensitivities.
As discussed above, this limit is in the range of interest for probing
new physics. Therefore it is interesting for future studies to
quantify the experimental errors which would result 
from larger data samples than the modest one assumed above.

\begin{table}
\caption{\small
Subset of results from the global form factor analysis
described in the text. The upper and lower limits of the couplings in their
departures from the SM values are given at 
68\% and 90\% CL. All couplings, each
with real and imaginary parts, can be determined in this way. The right-handed
charged-current coupling is shown both for unpolarized and 80\% left-polarized
electron beam, whereas the other results assume 80\% left-polarized 
beam only. $\Im$ is the imaginary part, otherwise the results listed here
are for the real parts. }
\begin{center}
\begin{tabular}[c]{ c | c | c | c}
Form Factor         & SM Value        & Limit       & Limit     \\
                    & (Lowest Order)  & 68\% CL     & 90\% CL   \\ 
\hline
$F_{1R}^W (P=0$)    & 0 &  $\pm 0.13$ &  $\pm 0.18$     \\
$F_{1R}^W (P=80\%$) & 0 &  $\pm 0.06$ &  $\pm 0.10$  \\
$F_{1A}^Z$          & 1 & $1\pm 0.08$ & $1\pm 0.13$              \\
$F_{1V}^Z$          & 1 & $1\pm 0.10$ & $1\pm 0.16$              \\
$F_{2A}^\gamma$     & 0 &  $\pm 0.05$ &  $\pm 0.08$           \\
$F_{2V}^\gamma$     & 0 &  $\pm 0.07$ &  $^{+0.13}_{-0.11}$   \\
$F_{2A}^Z$          & 0 &  $\pm 0.09$ &  $\pm 0.15$                \\
$F_{2V}^Z$          & 0 &  $\pm 0.07$ &  $\pm 0.10$                \\
$\Im (F_{2A}^Z)$    & 0 &  $\pm 0.06$ &  $\pm 0.09$          \\
\end{tabular}
\end{center}
\label{table:top-couplings}
\end{table}

\subsection{Helicity Analysis at Hadron Colliders}
\label{sec:helicity-hadron}

As discussed above, the Standard Model makes the firm prediction that
the $W$ polarization in top decays depends only on $m_t$ and $M_W$.
For $m_t=175$~GeV, the fraction of longitudinally polarized $W$'s in
top decay is roughly 70\%, with the remaining $W$'s being left-hand polarized.
This prediction, which is a direct consequence of the Lorentz structure of the
$t$-$W$-$b$ vertex,
can be tested in the large $\ttbar$ samples expected at
the Tevatron and LHC. Non-universal top couplings may manifest themselves
in a departure of $B(t\rightarrow bW_{\rm long})$ from its expected value.

The $W$ polarization can be measured in lepton + jets final states by
analyzing the angular distribution of the charged lepton from the decay
$t\rightarrow Wb$ followed by $W\rightarrow l\nu$.
The polarization of the $W$ is related to the charged lepton helicity
angle $\theta^*_l$, which is defined to be the emission angle of the
lepton in the rest frame of the $W$, with respect to the direction of the $W$ 
in the rest frame of the top. (It is equivalent to the angle $\chi_W$ of
Fig.~\ref{fig:helicity-angles}.)
This angle can be expressed
in terms of quantities measured in the lab frame via\cite{KaneYuanLad}
\begin{equation}
\cos\theta^*_l \approx \frac{2m^2_{lb}}{m^2_{l\nu b} - M_W^2} -1.
\end{equation}
Here $m_{lb}$ is the invariant mass of the charged lepton and the $b$,
and $M_{l\nu b}$ is the invariant mass of the lepton, the $b$, and the
neutrino, nominally equal to $m_t$. 

\begin{figure}[htbp]
\begin{center}
\leavevmode
\epsfysize=2.9in
\epsffile[0 162 500 657]{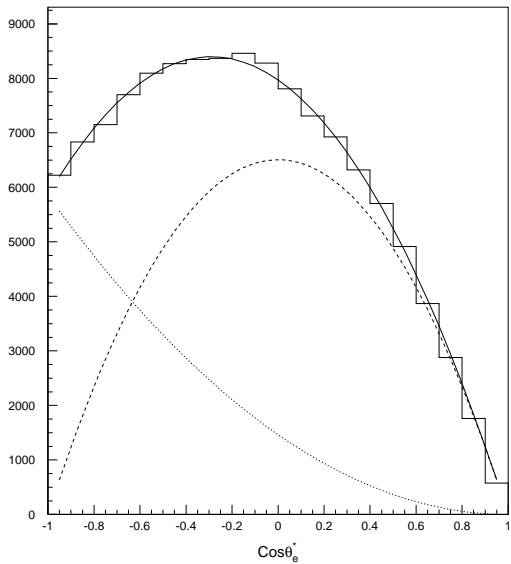}
\caption{The parton-level $\cos\theta^*_l$ distribution for $m_t = 170$~GeV.
The contributions from left-handed and longitudinally polarized $W$'s are
shown as the dotted and dashed lines respectively.}
\label{fig:winn_plot1}
\end{center}
\end{figure}

\begin{figure}[htbp]
\begin{center}
\leavevmode
\epsfysize=2.9in
\epsffile[0 162 500 657]{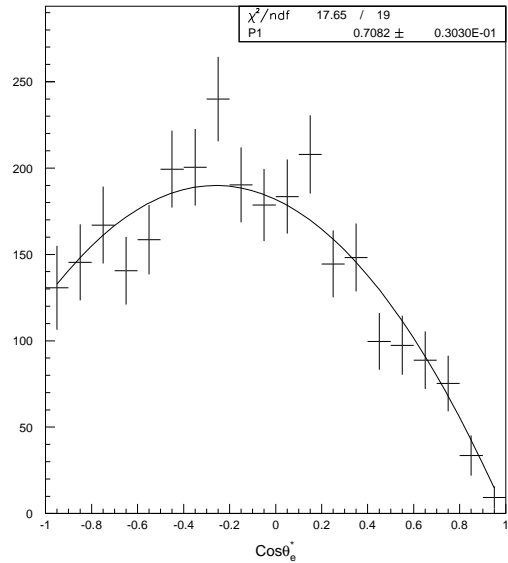}
\caption{Parton-level, acceptance-corrected $\cos\theta^*_l$ distribution for 1000 events, together with a fit to the Standard Model hypothesis.}
\label{fig:winn_plot4}
\end{center}
\end{figure}

The experimental strategy is to use the constrained fit described in
Section~\ref{sec:mass-hadron} to obtain the jet-parton correspondence,
which allows one to evaluate the invariant mass combinations. The
resulting $\cos\theta^*_l$ distribution is then fitted to a superposition of
$W$ helicity amplitudes in order to extract the fractions of $W_{\rm left}$,
$W_{\rm long}$, and $W_{\rm right}$, which contribute to $\cos\theta^*_l$ like 
$\frac{1}{4}(1-\cos\theta^*_l)^2$, $\frac{1}{2}sin^2\theta^*_l$, and
$\frac{1}{4}(1+\cos\theta^*_l)^2$ respectively.
A model analysis of this type at the Tevatron has been performed by
Winn\cite{TeV2000}. The $\cos\theta^*_l$ distribution
at the parton level, assuming
perfect resolution and no combinatoric misassignments, is shown in 
Fig.~\ref{fig:winn_plot1}. Note that a right-handed component would
peak near $\cos\theta^*_l = 1$, where the Standard Model predicts few
events. 
To determine best-case statistical precision of this measurement, Monte
Carlo\cite{Yuan-MC} pseudo-experiments are performed with top samples of 
various sizes,
still assuming perfect resolution and jet-parton assignment, but correcting
the acceptance with a $\cos\theta^*_l$-dependent factor. A fit to a 
sample of 1000 events is shown in Fig.~\ref{fig:winn_plot4}. 
The fit
accurately returns the input longitudinal $W$ fraction of 69\% to within
a 3\% statistical uncertainty. The statistical uncertainty in this best-case
scenario is found to behave like $1/\sqrt{N}$.

In a real experiment the precision will be lower due to the same effects
that complicate the mass measurement: combinatoric misassignment of the top
decay products, detector resolution, and backgrounds. The impact of
these effects on the helicity analysis has not yet been evaluated in 
detail. However, since this analysis uses the same constrained fit as 
the mass measurement, it is reasonable to assume that these effects would
be of the same order of magnitude in both analyses. In the mass analysis,
these effects lead to a degradation in resolution that is approximately
equivalent to a reduction in statistics by a factor of two, i.e. a
reduction in precision by a factor of $\sqrt{2}$. If this holds true
for the helicity analysis as well, then with a 10~fb$^{-1}$ sample at the
Tevatron it would be possible to measure the branching fraction to 
$W_{\rm long}$ to approximately 2\%, and to have sensitivity to a right-handed
component at the $\approx$1\% level. 

Neutral-current electroweak couplings of the top quark are not accessible 
at hadron colliders due to the dominance of strong production mechanisms
(or, in the case of single top, production through the weak charged current).
Final-state couplings of the top to the photon and $Z$ are extremely small.
A study of the neutral-current couplings is therefore the domain of 
$l^+l^-$ colliders.

\section{The $\lowercase{tt}H$ Coupling}
\label{sec:ttH}
     The role that the large top mass plays in electroweak symmetry
breaking can be directly explored by measuring the top-Higgs Yukawa
coupling.  In the Standard Model, this coupling strength $\lambda_t$
is proportional to the top mass: $\lambda_t=2^{1/4} \sqrt{G_F} m_t$.  The
top-Higgs coupling is consequently large and can be directly measured.
Such measurements are possible at both the LHC and the NLC.
The measurements are challenging in both environments, requiring
design-level luminosities for adequate statistics.

     The process $pp\rightarrow\ttbar H +X$ has been studied at the 
LHC for Higgs masses up to 120 GeV\cite{ATLAS}.  
The process relies on the availability of good vertex
detection even at the highest LHC luminosities for efficient $b$-tagging.
The Higgs is identified as a bump over a large background in the $b b$
invariant mass distribution
in events with a trigger lepton and at least three $b$ jets.  
The dominant backgrounds are due to $\ttbar$ and $W$ production with 
additional jets, some
of which are misidentified as $b$ jets. With 100~fb$^{-1}$, signals of more
than $3\sigma$ significance are expected for $m_H < 115$~GeV.  In principle
this signal could yield a measurement of the top-Higgs coupling, but
no such analysis is discussed.

    Several techniques can be applied at the NLC.  The top coupling
to a light Higgs ($m_H<2M_W$) can be measured at a 500 GeV collider with 
accurate
cross-section measurements  at $\ttbar$ threshold or by measuring the rate
of $\ttbar H$ events at $\sqrt{s}=500$~GeV.
Higher energies ($\sqrt{s}=1$ or 1.5 TeV) are needed to study the coupling
for intermediate or high Higgs mass ($m_H>2m_t$).

     The presence of an additional attractive force arising from Higgs
exchange produces a distinctive distortion in the cross section for
$\ttbar$ production near the 1S resonance.  This was discussed in the 
Section~\ref{sec:thresh-yukawa} above.  The size of the distortion is 
proportional to $\lambda_t^2/m_H$.  
The coupling could be measured to at least 10\% for
$m_H=100$~GeV with a 100 fb$^{-1}$ threshold scan.

     The yield of $\ttbar H$ events is proportional to the square of the
top-Higgs coupling.  The cross section for the process is small, of order
1~fb at a 500 GeV NLC for $m_H=100$~GeV; it grows to a few fb by 
1~TeV\cite{ttH:xsec}.
The final state typically contains eight jets, including four $b$ jets.  
Preliminary
studies\cite{Fujii-1,Euro-LCWS95} indicate that $\ttbar Z$ 
and $\ttbar j j$ events are significant
backgrounds. The top-Higgs coupling could be measured to 25\% with 
100~fb$^{-1}$
if $m_H=100$~GeV at $\sqrt{s}=500$~GeV.  The accuracy and Higgs mass reach
improve at higher energies. For example, Fujii finds a 10\% measurement
is possible at $\sqrt{s}=700$~GeV with the same integrated luminosity.
Studies are needed to quantify sensitivity to  intermediate and high
mass Higgs at higher $\sqrt{s}$.

The Higgs-strahlung process ($\ttbar H$) is also 
sensitive to effects that might
arise from extended Higgs sectors.  The interference between Higgs emission
from a virtual $Z$ and Higgs-strahlung from the final $t$ quark gives rise
to CP-violating effects in two Higgs doublet models.  This was studied
in Ref.~\cite{ttH:CP} where it was found that CP-violation effects could be 
seen at
$3\sigma$ level with several hundred $\ttbar H$ events and the most favorable
parameter choices.  Such studies will require center of mass energies
above 800~GeV and integrated luminosities of 300 fb$^{-1}$ or more.
Gunion and He presented\cite{Gunion} an analysis to discriminate 
between different models of the Higgs sector, using
two-Higgs-doublet models to exemplify the technique, which consists of 
measurements of the $\ttbar h$ differential cross section together with the
$Zh$ total cross section, where $h$ is a neutral Higgs boson.
For $m_h=100$ GeV, $\sqrt{s}=1000$ GeV, 
and an integrated luminosity of 500 fb$^{-1}$, they find that the Yukawa couplings
and Higgs model can be accurately determined.

     Measuring the coupling of the top quark to a heavy Higgs ($m_H>2m_t$) 
requires high center of mass energies and high intergrated luminosity.  
Three processes are of interest: $e^+e^-\rightarrow \ttbar H$; 
$e^+e^-\rightarrow \ttbar Z$; and $e^+e^-\rightarrow \nu \bar{\nu}\ttbar$.
Only the latter two have been studied.

     The cross-section for $e^+e^-\rightarrow\ttbar Z$ is about 5 fb between 
500 and 1000
GeV.  It is enhanced by the process $e^+e^-\rightarrow Z H$ when the 
Higgs subsequently
decays to $\ttbar$. For $m_H=500$~GeV, the enhancement is about 2~fb at 
$\sqrt{s}=1000$~GeV.  Fujii \textit{et al.}\cite{Fujii-2}, have studied 
this process.  
They enrich their
Higgs sample by first requiring a $\ttbar Z$ final state, and then cutting
on the appropriate $\ttbar$ invariant mass.  Extrapolating their results to
$\sqrt{s}=1000$~GeV and assuming $m_H=500$~GeV, leads to an estimated precision
in the top-Higgs coupling of 20\% for a 100 fb$^{-1}$ data set.

     Higgs enhancements are more dramatic in the reaction  $e^+e^-\rightarrow
\nu \bar{\nu} \ttbar$.  At $\sqrt{s}=1500$~GeV, the cross-section for this 
process is about 2~fb
in the absence of a Higgs, but will be enhanced by more than a factor of
two for Higgs masses in the range 400 to 850 GeV.  Peak sensitivities,
which occur when $m_H=500$~GeV, are nearly 10 times the nominal rate.
Preliminary studies by Fujii\cite{Fujii-2} show that care is required 
to eliminate
radiative $\ttbar$, $e^+e^-\ttbar$, and $\ttbar Z$ backgrounds.  They find 
that the top Higgs coupling can be measured to 10\% with 300 fb$^{-1}$ at 
$\sqrt{s}$ of 1000 GeV for $m_H=600$~GeV.

\section{Rare and Nonstandard Decays}

     The search for and discovery of the top quark at Fermilab has
relied on the assumption that the standard model decay $t\rightarrow Wb$ 
dominates.  This fact
is far from established, of course.  In fact,  the interesting 
speculation\cite{Kane-gluino}
that a conspiracy of SUSY-enhanced production balancing SUSY-depleted
decays explains the observed $t\bar{t}$ signal has not been excluded as yet.
The top width is unknown, and present estimates of the branching ratio 
$t\rightarrow Wb$
are model dependent; so there are only weak experimental constraints on
non-standard top decays.  The high top mass opens the kinematic window for
decays to new, massive states, such as those inspired by supersymmetric
models, $t\rightarrow\tilde{t}+$ neutralino ($\chi^0$) and $t\rightarrow H^+b$.  The 
high top mass also encourages
speculation that neutral-current decays, like $t\rightarrow c \gamma$ 
or $t\rightarrow c Z$, may
be large enough to be interesting experimentally.
     If the stop and neutralino masses are low enough, the decay $t\rightarrow
\tilde{t}\chi^0$ can occur with a sizable branching fraction.
Typically, one imagines that the neutralino escapes undetected and that
the subsequent decay, $\tilde{t}\rightarrow c\chi^0$, leaves a lone remnant 
hadronic
jet and missing energy.  It is reasonable to expect that this issue
will be addressed with present and future Fermilab data
by searching for events
with an identified $t$, a charm jet, and missing energy.  Venturi\cite{Venturi}
has studied how to detect this decay at an NLC, which is done by looking
for an event where the invariant mass of one hemisphere is near the top
mass, and the other is substantially below.  He finds that a 10~fb$^{-1}$ data
set is sufficient to establish a 3$\sigma$ discovery, provided the branching
fraction is $> 2\%$ (for $m_{\tilde{t}}=80$ GeV and $m{\chi^0}=55$ GeV).

     Top decays to a charged Higgs, $t\rightarrow H^+b$, are also expected in 
supersymmetric models when the decay is kinematically allowed.
The charged Higgs is expected to decay predominantly to $\tau \nu_{\tau}$
when $\tan\beta>1$, so the appropriate signature is an apparent violation of
lepton universality in top decays, leading to an excess of taus in the
top decay products.  Run 2 at the Tevatron will be sensitive to branching
fractions $B(t\rightarrow H^+b)> 11\%$\cite{TeV2000}.  At LHC, the decay is 
detectable if 
$m_H<130$~GeV for most values of $\tan\beta$ with 
10~fb$^{-1}$\cite{ATLAS}.  At NLC a study\cite{Venturi}
has shown that the decay is observable if $m_H<125$~GeV, essentially
independent of the value of $\tan\beta$, with 100~fb$^{-1}$.

     The FCNC decays $t\rightarrow c \gamma$ 
and $t\rightarrow c Z$ are tremendously suppressed
in the Standard Model, with branching fractions of order $10^{-12}$.  
Consequently
their observation at detectable levels is a robust indication of new physics.
Models with singlet quarks or compositeness could have branching ratios
for these decays as large as 1\%\cite{Han-Hewett}.  The signature for 
these decays, a very
high $P_T$ photon or a high energy lepton pair with an invariant mass 
consistent
with the $Z$ mass, are distinctive enough to permit sensitive searches in the
hadronic environment.  Run 2 at the Tevatron will probe to branching 
fractions of about $3\times 10^{-3}$ ($2 \times 10^{-2}$) for 
$t\rightarrow c \gamma$ ($t\rightarrow c Z$)\cite{TeV2000}.  At the LHC
with its very large top samples,  branching fractions as small as $5 \times 
10^{-5}$
could be measured for $t\rightarrow c Z$, assuming an integrated luminosity 
of 100~fb$^{-1}$\cite{ATLAS}.
At NLC, the sensitivity is limited by the available statistics to of order 
10$^{-4}$ for $t\rightarrow c \gamma$ 
and 10$^{-3}$ for $t\rightarrow c Z$, assuming an integrated luminosity
of 50~fb$^{-1}$.  Similar limits could be established by looking directly for
$e^+e^-\rightarrow t\bar{c}$ events\cite{Han-Hewett}.

\section{Conclusions}  

The systematic study of the top quark offers many possibilities for exploring
physics beyond the standard model. Because the top quark mass enters 
quadratically into the $\rho$-parameter, a precision 
measurement of $m_t$ can be used together with $M_W$ to constrain the Higgs
mass. In the exciting event that a Higgs particle is observed, knowledge of
$m_t$ will help determine whether it is a standard model Higgs or some other,
more exotic variety. Measurements of the top couplings and form factors 
directly probe the weak interactions of a bare quark at their natural
scale, and anomalies in these couplings could signal the presence of new
physics at the TeV scale or higher. Direct measurements of the top width
and $V_{tb}$ could reveal the existence of nonstandard decay modes or 
additional quark generations. And the top-Higgs Yukawa coupling can be
probed directly, particularly if the Higgs is light. Each of these measurements
is of great interest and should play an important role in planning future
experiments. 

The Fermilab Tevatron will be the only facility capable of studying
the top quark until the LHC turns on in 2005. With 30~fb$^{-1}$ delivered
in ``Run III'' following the initial Main Injector collider run, a 
top mass uncertainty of $\leapp 2$~GeV appears feasible. This measurement would
be sufficiently accurate that uncertainties in other quantities ($M_W$,
$\sin^2\theta_W$,$\alpha_s(M_Z^2)$) would dominate the precision electroweak
fits. The Tevatron can measure $\Gamma_t$ and $V_{tb}$ to better than
10\%, albeit with some model-dependent assumptions. The Tevatron will also test
the charged-current form-factors and search for rare and nonstandard decays.
Its main advantage, of course, is that it exists and has a monopoly on the
subject for roughly the next decade. The Tevatron program should take full
advantage of this situation and maximize the integrated luminosity before
the LHC turn-on.

The LHC, with its enormous top production cross section, is a veritable top
factory. In particular, its sensitivity to rare decays is unlikely to be
matched by other machines. As is the case for the Tevatron, many 
measurements will be systematics-limited. Neither LHC experiment, for example,
is currently willing to claim a mass measurement better than 2~GeV. However,
the very large control samples that will be available at the LHC
suggest that these systematics might be better controlled, or that precision
measurements could be performed using small, very clean subsamples.
The measurement of the top-Higgs coupling at the LHC will be extremely
challenging due to the low cross section and difficult backgrounds. In 
general, top physics at the LHC has not been studied in the same level
of detail as, say, Higgs and SUSY searches. It could benefit from additional
study since its potential has not been fully explored.

An $e^+e^-$ linear collider offers the greatest potential for high-precision
top physics in the LHC era. If the beam energy spectrum can be understood
to the level expected, the top mass can be measured to better than 200~MeV.
A number of fundamental parameters can be measured at the $t\bar{t}$ threshold,
including $\Gamma_t$, $V_{tb}$, $\alpha_s$, the charge and spin of the
top quark, and the top-Higgs Yukawa coupling if the Higgs is sufficiently
light. The full array of top gauge couplings can be measured, including
the neutral-current couplings, which are inaccessible at hadron colliders. 
The top-Higgs coupling can be measured in the open top region as well,
though this will require extended running at design luminosity. If the
Higgs (or \textit{a} Higgs) is light enough for this measurement to be made,
it will also be light enough to have been directly observed at the NLC, LHC,
or even perhaps the Tevatron. The Yukawa coupling of this particle
to the top quark may depend on whether it is a standard model Higgs, a
SUSY Higgs, or some other thing entirely. A direct measurement of this
coupling will thus address head-on the question of how the top quark,
and by extension all fermions, acquire mass.


\begin{thebibliography}{99}
%
%
\bibitem{cdf_obs} F.~Abe \etal~(CDF Collaboration), \prl{74},2626 (1995).
\bibitem{d0_obs} S.~Abachi \etal~(D0 Collaboration), \prl{74}, 2632 (1995).
\bibitem{DG_snowmass} For a review of the top mass and other measurements,
     see D.~Gerdes, ``Top Quark Physics Results from CDF and D0,'' these
     proceedings; hep-ex/9609013 (1996).
\bibitem{TeV2000} ``Future Electroweak Physics at the Fermilab Tevatron:
    Report of the TEV2000 Study Group,'' D.~Amidei and R.~Brock eds.,
    Fermilab-Pub-96/082 (1996).
\bibitem{Peskin-Murayama} M.~E.~Peskin and H.~Murayama, ``Physics Opportunities
    of $e^+e^-$ Linear Colliders,'' Ann. Rev. Nucl. Part. Sci.
    \textbf{46}, 533 (1996); hep-ex/9606003 (1996).
\bibitem{Frey96} R.~Frey, ``Top Quark Physics at a Future $e^+e^-$ Linear
    Collider: Experimental Aspects,'' proceedings of the Workshop on
    Physics and Experiments with Linear Colliders
    (LCWS95), Iwate, Japan, Sept., 1995; hep-ph/9606201 (1996).
%
%
\bibitem{xsec_theory} E.~L.~Berger and H.~Contapaganos, \prd{54}, 3085 (1996);
    S.~Catani, M.~L.~Mangano, P.~Nason, and L.~Trentadue, Phys. Lett.
    \textbf{B378}, 329 (1996); S.~Catani, M.~L.~Mangano, P.~Nason, and
    L.~Trentadue, CERN-TH/96-86, hep-ph/9604351 (1996).
    E.~Laenen, J.~Smith, and
    W.~L.~Van Neerven, Phys. Lett. \textbf{B321}, 254 (1994).
\bibitem{single} S.~Dawson, Nucl. Phys. \textbf{B249}, 42 (1985); 
    S.~Willenbrock and D.~Dicus, \prd{34}, 155 (1986);
    S.~Dawson and S.~Willenbrock, Nucl. Phys. \textbf{B284}, 449 (1987);
    C.-P.~Yuan, \prd{41}, 42 (1990); S.~Cortese and R.~Petronzio,
    Phys. Lett. \textbf{B253}, 494 (1991); G.~V. Jikia and S.~R.~Slabospitsky,
    Phys. Lett. \textbf{B295}, 136 (1992); R.~K.~Ellis and S.~Parke, 
    \prd{46}, 3785 (1992); G.~Bordes and B.~van Eijk, Z. Phys. \textbf{C57},
    81 (1993); G.~Bordes and B.~van Eijk, Nucl. Phys \textbf{B435}, 23 (1995);
    D.~O.~Carlson and C.-P.~Yuan, Phys. Lett \textbf{B306}, 386 (1993);
    T.~Stelzer and S.~Willenbrock, Phys. Lett. \textbf{B357}, 125 (1995).
%
%
\bibitem{Barklow} T.~Barklow, talk presented at this meeting.
\bibitem{Fujii-1} K.~Fujii, T.~Matsui, and Y.~Sumino, Phys. Rev. 
    \textbf{D50}, 4341 (1994).
\bibitem{Jackson} D.J.~Jackson, talk presented at this meeting.
%
%
\bibitem{peskin} M. Peskin and T. Takeuchi, Phys. Rev. D 46,381 (1992)
\bibitem{sarid} U. Sarid, these proceedings; hep-ph/9610341 (1996).
\bibitem{tipton} P. Tipton for CDF Collaboration, XXVIII International
Conference on High Energy Physics, Warsaw, Poland, July, 1996.
\bibitem{strovink} M. Strovink, ``The D0 Top Quark Mass Analysis,'' to appear
    in the proceedings of the 11th Topical Workshop on Proton-Antiproton
    Collider Physics, Padua, Abano Terme, Italy, 26 May - 1 June 1996;
     FERMILAB-CONF-96/336-E (1996).
\bibitem{heinson} A. Heinson ``Future Top Physics
at the Tevatron and LHC,'' Proceedings of the XXXIst Rencontres
de Moriond, QCD and High Energy Hadronic Interactions, Les Arcs,
Savoie, France, 23rd-30th March 1996.

\bibitem{ATLAS} ATLAS Technical Proposal, CERN/LHCC 94-43, LHCC/P2 (1994).
\bibitem{raja} R. Raja, these proceedings; hep-ex/9609016 (1996).
\bibitem{orr} L. H. Orr, T. Stelzer, and W.J. Stirling, Phys. Rev. 
     \textbf{D52}, 124 (1995).
%
%
\bibitem{Peskin} M.~Strassler and M.~Peskin, 
Phys. Rev. \textbf{D43}, 1500 (1991).
%
\bibitem{Fadin-1} V.~Fadin and V.~Khoze, 
JETP Lett. \textbf{46}, 525 (1987) and
Sov. J. Nucl. Phys. \textbf{48}, 309 (1988).
\bibitem{Jezabek-1} M.~Jezabek, J.~Kuhn, and T.~Teubner, 
Z. Phys. \textbf{C56}, 653 (1992).
\bibitem{Jezabek-2}
M.~Jezabek and T.~Teubner, Z. Phys. \textbf{C59}, 669 (1993).
\bibitem{Sumino-1} Y.~Sumino, K.~Fujii, K.~Hagiwara, H.~Murayama, and C.-K.~Ng,
Phys. Rev. \textbf{D47}, 56 (1993).
\bibitem{Fadin-2} E.A.~Kuraev and V.S.~Fadin, Sov. J. Nucl. Phys.
\textbf{41}, 466 (1985).
\bibitem{Chen} Pisin Chen, Phys. Rev. \textbf{D46}, 1186 (1992).
\bibitem{ZDR-IR} C.~Adolphsen, {\it et al.,} ``Zeroth-Order 
Design for the NLC'', Ch. 12, SLAC Report 474, May 1996;
and references therein.
\bibitem{Hawaii} P.~Igo-Kemenes, M.~Martinez, R.~Miquel, and S.~Orteu,
proceedings of the Workshop on Physics and 
Experiments with Linear Colliders (LCWS93), Waikoloa, Hawaii, USA, 1993.
\bibitem{Fujii-2}  K. Fujii, proceedings of the 1995 
SLAC Summer Institute.
\bibitem{Euro-LCWS95} P.~Comas, R.~Miquel, M.~Martinez, and S.~Orteu, 
``Recent Studies on Top Quark Physics at NLC'', proceedings of 
 the Workshop on Physics and Experiments with Linear Colliders (LCWS95), 1995.
\bibitem{Kuhn-Yukawa} R.~Harlander, M.~Jezabek, and J.H.~Kuhn,
Acta. Phys. Polon. \textbf{27}, 1781 (1996), hep-ph/9506292 (1995).
\bibitem{Sumino-2}H.~Murayama and Y.~Sumino, Phys. Rev. \textbf{D47}, 82 (1993).
\bibitem{Miller} N.M. Frary and D. Miller, DESY 92-123A, Vol. I, 1992,
p. 379.
%
%
\bibitem{stelzer} T.~Stelzer and S.~Willenbrock, Phys. Lett. \textbf{B357},
    125 (1995).
\bibitem{stelzer_snowmass} T.~Stelzer, these proceedings.
%
%

%
%
\bibitem{Suzuki} W.~Bernreuther and M.~Suzuki, Rev. Mod. Phys. 
\textbf{63}, 313 (1991).
\bibitem{Bernreuther} W.~Bernreuther, T.~Schrooder, and T.N. Pham,
Phys. Lett. \textbf{B279}, 389 (1992).
\bibitem{Soni-1} A.~Soni and R.~Xu, Phys. Rev. Lett. \textbf{69}, 33 (1992).
\bibitem{Rizzo} T.G.~Rizzo, these proceedings; hep-ph/9610373 (1996).
%
%
\bibitem{Yuan}  G.A.~Ladinsky and C.P.~Yuan, Phys.
Rev. \textbf{D49}, 4415 (1994); see also references therein.
\bibitem{Peskin-2} M.E.~Peskin and C.R.~Schmidt, proceedings of the 
Workshop on Physics and Experiments
with Linear Colliders (LCWS91), Saariselka, Finland, 1991.
\bibitem{Schmidt} C.R.~Schmidt, SCIPP-95/14 (1995); hep-ph/9504434 (1995).
\bibitem{Parke} S.~Parke and Y.~Shadmi, Phys. Lett. \textbf{B387}, 199 (1996).
\bibitem{Fero} M.~Fero, talk presented at this meeting.
\bibitem{Soni-2} D.~Atwood and A.~Soni, Phys. Rev. \textbf{D45}, 2405 (1992).
\bibitem{Cuypers} F.~Cuypers, proceedings of the Workshop on Physics and Experiments with Linear Colliders (LCWS95), 1995.

%
\bibitem{KaneYuanLad} G.~L.~Kane, C.~P.~Yuan, and G.~Ladinsky,
Phys. Rev. \textbf{D45}, 124 (1992).
\bibitem{Yuan-MC} The event generator used is that of E.~Malkawi and
C.~P.~Yuan, Phys. Rev. \textbf{D50}, 4462 (1994).
%
%
\bibitem{ttH:xsec} A.Djouadi, J. Kalinowski and P.M. Zerwas, 
    Z. Phys. \textbf{C54}, 255 (1992).
\bibitem{ttH:CP} S. Bar-Shalom, D. Atwood, G. Eilam, R. Mendel, and A. Soni, 
    Phys. Rev. \textbf{D53}, 1162 (1996).
\bibitem{Gunion} J.F.~Gunion and X.-G.~He, these proceedings; 
hep-ph/9609453 (1996).
%
%
\bibitem{Kane-gluino} G.~L.~Kane and S. Mrenna, Phys. Rev. Lett \textbf{77},
    3502 (1996).
\bibitem{Venturi} A. Venturi, in proceedings of the Workshop 
on Physics and Experiments 
with Linear Colliders (LCWS93), Waikoloa, Hawaii, USA, 1993.
\bibitem{Han-Hewett} T.~Han and J.~Hewett, SLAC-PUB-7178, Dec. 1996.

\end{thebibliography}
\end{document}